# LaAlO$_3$/SrTiO$_3$: a tale of two magnetisms

Yun-Yi Pai[1,2], Anthony Tylan-Tyler[1,2], Patrick Irvin[1,2] and Jeremy Levy[1,2]

[1]Department of Physics and Astronomy, University of Pittsburgh, Pittsburgh, PA 15260 USA

[2]Pittsburgh Quantum Institute, Pittsburgh, PA 15260 USA

**Introduction**

Complex oxides have been a continual source of discoveries in solid state physics. Recent advances in thin-film growth techniques have enable unit-cell-level control over their composition, resulting in the creation of layered, complex oxide heterostructures. *Emergent phenomena*—properties or phases observed only within heterostructures and not found in the parent compounds—provide a recurring theme for this active interdisciplinary field that spans materials science, physics, chemistry, and engineering. One of the most striking, and controversial, examples of an emergent property is the observation of magnetism at the interface formed between the two complex oxides LaAlO$_3$ and SrTiO$_3$. Both materials are separately non-magnetic, and yet there are many reports of magnetic phenomena associated with the LaAlO$_3$/SrTiO$_3$ interface.[1]

The properties of SrTiO$_3$ have fascinated researchers for generations. This simple perovskite material has been widely utilized as a substrate for the growth of other materials, but has its own wide-ranging properties that span "all of solid state physics"—*save for magnetism*

---

[1] It should be noted that there are also many complex-oxide heterostructures with magnetic properties that are derived from bulk behavior. For a more general discussion of magnetism at complex oxide interfaces, readers are referred to the chapter by Dagotto et al.

[1] (see Figure 1). SrTiO$_3$ has also inspired important discoveries such as high-temperature superconductivity [2]. Interest in the fundamental properties of SrTiO$_3$ was renewed in the last decade due to the 2004 report by Ohtomo and Hwang of high mobility electron transport at the interface between LaAlO$_3$ and TiO$_2$-terminated SrTiO$_3$ [3]. The discovery of magnetism at the LaAlO$_3$/SrTiO$_3$ interface was first reported by Brinkman et al. [4] in 2007. Following that initial report, a cascade of new signatures of magnetism at the LaAlO$_3$/SrTiO$_3$ interface were obtained using a variety of measurement techniques.

While most published findings are internally consistent, some reports apparently contradict the existence of interfacial magnetism, and some magnetic signatures have not always been reproduced under nominally identical conditions. The goal of this chapter is to help organize and classify the, sometimes contradictory, observations that have been made on LaAlO$_3$/SrTiO$_3$ heterostructures, and closely related systems, which have been investigated extensively as well as explored theoretically. We argue that there is not a single form of magnetism but, rather, that there are two main types of magnetic behavior with distinct properties and origins.

## *Organization of this chapter*

The purpose of this chapter is twofold. First, we will review the experimental evidence for magnetic phenomena at the LaAlO$_3$/SrTiO$_3$ interface. We will then argue that essentially all of the signatures of magnetism can be sorted into two distinct categories: (1) magnetic phases (e.g., ferromagnetic or Kondo) involving local magnetic moments and their coupling to itinerant electrons; (2) *metamagnetic* effects that are mediated by attractive electron-electron interactions that do not involve local moments. We will review possible candidates for the local moments that give rise to the ferromagnetic phases and focus on arguments for one potential source: oxygen vacancies. For the metamagnetic transport signatures, band-structure effects (e.g., Lifshitz transition) and strong attractive electron-electron interaction can help consolidate disparate experimental findings. The coexistence of magnetism and superconductivity is discussed briefly.

While the main focus will be on experimental evidence, we will also briefly summarize various theoretical approaches that have been taken so far. However, we argue that the starting point (assumptions) for many theories could be hindered by attempts to combine these two distinct (and weakly interacting) forms of magnetism. To make better progress theoretically, we argue that it will be simpler to initially restrict the domain of inquiry to one or the other phases. Toward the end of the article, we discuss open questions as well as some tests for this framework for understanding the two classes of magnetic phenomena.

**Experimental Signatures of Magnetism**

The first signature of magnetism at the $LaAlO_3/SrTiO_3$ interface was reported in 2007 by Brinkman et al. [4]. They reported a hysteresis loop in the magnetoresistance (Figure 7a) and a characteristic Kondo minimum (followed by a saturation at lower temperature) in the resistance versus temperature that was attributed to magnetic impurities (Figure 2a). The Kondo signature was later also observed by Hu et al. [5], as well as by Lee et al. [6] in electrolyte-gated $SrTiO_3$. Several other signatures of magnetism followed, including observations of anisotropic magnetoresistance [5,7-13], anomalous Hall effect [11,14-16], and direct observation of magnetization with tools from cantilever magnetometry [17], scanning SQUID magnetometry and susceptometry [18-20], magnetic force microscopy [21,22], and x-ray magnetic circular dichroism [23,24]. Below, we consider these findings in more detail.

*Magnetometry*

*SQUID Magnetometry*

In 2011, Ariando et al. [9] reported a direct measurement of LaAlO$_3$/SrTiO$_3$ magnetization using SQUID magnetometry. As shown in Figure 2b, a series of hysteresis loops of the magnetization in external magnetic field is observed. This hysteresis persists up to room temperature. It is worth noting that this ferromagnetic hysteresis was only reported for the samples grown at P(O$_2$) = 10$^{-2}$ mbar pressures, which exist at the boundary of the metal-insulator transition. A similar finding for Nb-doped SrTiO$_3$ has also been reported by Liu et al. [25].

*Cantilever magnetometry*

Using cantilever-based torque magnetometry (Figure 2c), Li et al. [17] reported an in-plane magnetization of the 2DEG at LaAlO$_3$/SrTiO$_3$ with magnitude corresponding to ~0.3 µ$_B$ per interface unit cell (if all the magnetization is assigned to the interface). The magnetization was found to be independent of temperature up to 40 K and persists beyond 200 K. The same sample was also found to be superconducting below 120 mK.

*Scanning SQUID microscopy*

Scanning SQUID microscopy has been used to image inhomogeneous magnetism in LaAlO$_3$/SrTiO$_3$. With a 3 µm-diameter SQUID loop serving as a local sensor of magnetic flux, this technique can detect magnetic moments as small as ~10$^2$ µ$_B$/$\sqrt{Hz}$ [26]. Bert et al. [18] reported dipole-like magnetic patches at the interface of LaAlO$_3$/SrTiO$_3$ (Figure 2d) [18]. Subsequent investigations by Kalisky et al. [19] determined that the patches were highly non-uniform in size, orientation, and physical placement, and only observed at or above a critical thickness for LaAlO$_3$ layer (3 u.c.), the same as the threshold for the insulator-to-metal transition in LaAlO$_3$/SrTiO$_3$ [27]. By touching the LaAlO$_3$ surface with the SQUID sensor, it was discovered that the magnetic moment and orientation of the ferromagnetic patches could be manipulated (Figure 11f) [20]. The same scanning SQUID system can also simultaneously image diamagnetic susceptibility associated with superconductivity [18,28]. Correlations between magnetic and superconducting order were not found within the spatial resolution of this technique.

*β-NMR*

Salman et al. [29] measured magnetic properties of LaAlO$_3$/SrTiO$_3$ using β-NMR. In this technique, spin-polarized radioactive $^8$Li atoms are shot at the LaAlO$_3$/SrTiO$_3$ sample. The spins of Li nuclei are inferred from the spins of electrons emitted via β decay. The presence of magnetic moments in the sample increases the spin decoherence rate. A "weak" ($\sim 1.8 \times 10^{-3} \mu_B/\text{u.c.}$) magnetization is obtained for both 6 and 8 u.c samples. If a high degree of spatial non-uniformity is assumed, then a local density $\sim 10^{12} \mu_B/\text{cm}^2$ may be present [29], which is on the same order of magnitude as Bert et al. [18].

*Magnetic force microscopy (MFM)*

Bi et al. [21] reported electronically-controlled magnetism at the interface of LaAlO$_3$/SrTiO$_3$ using magnetic force microscopy (Figure 2e). When an in-plane polarized magnetic tip in an atomic force microscope (AFM) is brought to the vicinity of the sample surface, the magnetization of the sample and its interaction with the MFM tip shifts the oscillation amplitude, phase, and frequency, which are recorded as a function of position. Magnetic interactions between the MFM tip and the interface take place through a thin layer of gold deposited on the surface of LaAlO$_3$ to serve as a top gate for the sample. The 2DEG at the interface can be tuned from insulating to conducting, which is monitored through the capacitance of the two-terminal device. Magnetic interactions are observed only when the tip is magnetized in-plane and when the device is gated into the insulating regime. Related measurements show a pronounced magnetoelectric effect which indicates that free carriers rapidly screen (on <10 μs time scales) ferromagnetic domains [21]. Follow-up measurements revealed that this electronically-controlled ferromagnetism is observed only within a thickness window 8-25 u.c. [22]. Outside of this region the samples cannot be gated into the insulating regime: samples with thinner LaAlO$_3$ layers are prone to leakage via direct tunneling, while samples with LaAlO$_3$ layer thicker than 25 u.c. are subjected to Zener tunneling.

*Transport*

*Anisotropic magnetoresistance*

Anisotropy in the magnetotransport measurements is a widely reported signature of magnetism at the LaAlO$_3$/SrTiO$_3$ interface [5,7-13]. When the magnetic field is (i) oriented out-of-plane, the magnetoresistance is positive [30]. When the magnetic field is (ii) in-plane and perpendicular to the direction of transport, the magnetoresistance is positive for small fields and eventually becomes negative at larger field [31]. When the magnetic field is (iii) parallel to the direction of transport, the magnetoresistance is negative [7,9]. As the applied magnetic field rotates in plane, as shown in Figure 3a, a sinusoidal dependence of magnetoresistance as a function of the angle φ between the field and the direction of current (Figure 3b). However, when the carrier concentration $n_e$ is above a critical value $n_L$, a Lifshitz transition occurs and this two-fold symmetry develops crystalline anisotropy with high harmonics (Figure 3c, e).

*Nonlinear and Anomalous Hall Effects*

Nonlinearity in the Hall coefficient is widely reported for SrTiO$_3$-based heterostructures [8,11,14-16,30,32-38]. Figure 5a [36] shows the most common nonlinearity in the Hall resistance $R_{xy}$, using a standard Hall geometry with an out-of-plane external magnetic field. This results in an overall parabolic shape of the Hall coefficient $R_H$ (the slope of the Hall resistance $R_{xy}$) in the nonlinear regime, as shown in Figure 5b [16]. This feature is commonly attributed to the multiband nature of SrTiO$_3$ [11,30,38].

The attribution of nonlinear Hall effect to the existence of multiple bands cannot explain the up-turn of $R_H$ at smaller fields. Gunkel et al. [16] postulated an additional anomalous Hall effect (AHE) with Langevin-type dependence on the external magnetic field [16]. While the physical origin for this AHE was not identified, it was considered to be a manifestation of some magnetizable component in the system, e.g., due to the release and alignment of the spin degree of freedom.

Using a different configuration, namely a large in-plane magnetic field with small out-of-plane component, as shown in the inset of Figure 5c, Joshua et al. [15] reported a relatively sharp onset of an anomalous Hall effect a critical in-plane field $H_c^{\parallel}$ that depends strongly on carrier concentration $n_e$. This critical field diverges near the Lifshitz density $n_L$: $H_c^{\parallel} \sim (n_e - n_L)^{-1}$, increasing from 2T to values as large as 14 T. This offset in Hall resistance was interpreted as some type of magnetization that has been "released" above a density-dependent critical magnetic field.

*Rashba spin-orbit coupling*

Strong gate-tunable Rashba spin-orbit coupling was first reported by Caviglia et al. [39] and Ben Shalom et al. [7]. Due to the polar nature of the LaAlO$_3$ layer, there is a built-in electric field in the direction perpendicular to the interface (along the c-axis for [001] the SrTiO$_3$). This field breaks inversion symmetry and introduces Rashba spin-orbit coupling (Figure 4) that mixes the $d_{xy}$ band and $d_{xz}/d_{yz}$ bands and spin-splitting. The spin-splitting is at its largest (several meV) near the avoided crossings of the bands.

Rashba spin-orbit coupling was probed via magneto-transport. As the gating voltage increases, Caviglia et al. found that the magnetoconductance changes from positive (associated with weak localization and negligible Rashba spin-orbit coupling) to negative (associated with weak anti-localization and strong Rashba spin-orbit coupling). By fitting the magnetoconductance to Maekawa-Fukuyama theory of spin-relaxation, the coupling strength was found to be strongly dependent on gate, with a sharp rise near the gating voltage corresponding to insulator to superconductor transition of the same sample.

Ben Shalom et al. [36], on the other hand, reported a field-independent longitudinal resistance for in-plane magnetic fields greater than the superconducting upper critical field $H_{c2}^{\parallel}$ but below a larger critical field $H^*$, followed by a significant drop of the resistance above $H^*$. Ben Shalom et al. [36] associated $H^*$ with a spin-orbit field, and obtain a dependence of Rashba spin-orbit coupling on gate voltage opposite to that reported by Caviglia et al [39].

A current-driven Rashba field was reported by Narayanapillai et al. [40]. Recently, using strong Rashba spin-orbit coupling at LaAlO$_3$/SrTiO$_3$ for spin-charge conversion has been demonstrated recently by Lesne et al. [41]. The reported spin-charge conversion efficiency is comparable to non-2D materials such as Pt, Ta, W [41]. Additionally, a sign change of the Rashba coefficient is also reported [41].

Attempts to resolve the spin-dependent band structure using spin-polarized ARPES (SARPES) have been made, with conflicting reports. A large spin-splitting of 90 meV was reported by Santander-Syro et al. on doped $SrTiO_3$ [42], followed by a null result reported by Walker et al. [43] and a possible reconciliation of the two contradictory reports [44].

*Electron pairing without superconductivity*

The superconducting state is associated with a variety of non-trivial magnetic properties. A conventional (type I) superconductor is a perfect diamagnet and the superconducting state consists of a condensate of Cooper pairs, which (in most cases) have a spin-singlet configuration. As will be described below, $LaAlO_3/SrTiO_3$ exhibits a robust phase in which electrons remain paired but are not superconducting. The pairing transition itself involves significant changes in magnetization and orbital character that are associated with the complex metamagnetic behavior observed in this system.

Superconductivity in $SrTiO_3$ was first reported in 1964 by Schooley et al [45]. Apart from being one of a few semiconductors to exhibit superconducting behavior, it was suspected that its nature might be different from conventional BCS superconductivity. For example, the superconducting "dome" observed for $SrTiO_3$ [46] bears a striking resemblance to high-temperature superconductors, which were discovered two decades later [2]. One mysterious and controversial phase of high-temperature superconductors is the so-called "pseudogap" regime in which superconductivity is not observed, but a gap in the single-particle spectrum remains as observed through tunneling experiments and other measurements [47]. While it is far from clear that $SrTiO_3$ shares a similar pairing mechanism (it lacks, for example, an antiferromagnetic parent phase that is believed to contribute to pairing in high-$T_C$ compounds), planar tunneling experiments on $LaAlO_3/SrTiO_3$ structures by Richter et al. [48] show a pseudogap feature past the boundary of the superconducting dome. The physical origin of the pseudogap phase in high-$T_c$ compounds continues to be debated. In particular, a seemingly straightforward question— whether the pseudogap phase is related to pre-formed Cooper pairs—has been challenging to resolve experimentally in the cuprates.

Cheng et al. [49] investigated the phenomenon of electron pairing using a single-electron transistor (SET) geometry (Figure 8a). A SET is a three-terminal device with a quantum dot (QD) that is tunnel-coupled to source and drain leads, with a gate that can change the number of carriers in the QD [50]. The conductance of a conventional SET is generally low if the energy of the lowest available state in the QD is not aligned with the chemical potential of the leads. When there is such an alignment, the energy of the state with *N* and *N*+1 electrons in the QD becomes degenerate, and electrons can tunnel resonantly through the device. Cheng et al. created a SET device using conductive-AFM (c-AFM) at the $LaAlO_3/SrTiO_3$ interface (Figure 8a). They found that the differential conductance of the SET (Figure 8b) departed significantly from what one would expect for a SET formed from ordinary semiconductors. Three distinct regimes of behavior were observed as a function of applied magnetic field (out of plane): (i) at very low magnetic fields (below $|B| < \mu_0 H_{c2} \sim 0.1$ T) a Josephson supercurrent (marked SC in Figure 8b) is found to flow through the device; (ii) at intermediate fields ($\mu_0 H_{c2} < |B| < B_P \sim 2 - 11$ T), a series of vertical (magnetic field-independent) lines are observed; (iii) $|B| > B_P$, the vertical lines bifurcate, leading to a doubling of the number of conductance peaks which Zeeman shift as the magnetic field is further increased. Significantly, while $B_p$ varies from device to device, it generally is found to increase as the carrier density decreases (Figure 12a). Furthermore, this value is more than an order of magnitude greater than the upper critical field for the superconductivity, $\mu_0 H_{C2}$ (Figure 12a), and for temperatures that greatly exceed the highest observed superconducting state in $SrTiO_3$. That is to say, it is a state in which there is *electron pairing without superconductivity*.

## *Coexistence of magnetism and superconductivity*

### *Superconductivity at LaAlO$_3$/SrTiO$_3$ interface*

The 2DEG exhibits superconductivity [51,52], at low carrier concentrations ($n_e \sim 10^{13}$ cm$^{-2}$) and critical temperatures in the range $T \approx 200 - 300$ mK [51]. Like most other characteristics of this interface, the superconducting properties can be tuned by a back gate [52]. By tracing out the critical temperature as a function of gating voltage (and therefore carrier concentration), a dome-shaped outline enclosing the superconducting phase is seen, similar to high-$T_c$ superconductors as well as the doped bulk SrTiO$_3$.

### *Magnetic effects*

While a true spatial coexistence of ferromagnetism and superconductivity at LaAlO$_3$/SrTiO$_3$ is not confirmed, coexistence at the sample level is reported by Li et al. [17] and Dikin et al. [53] and within the resolution of the scanning SQUID (~3 µm) by Bert et al. [18]. Dikin et al. [53] observed a clear hysteresis in magnetotransport measurements as a function of external magnetic field by mapping the critical temperature as a function of out-of-plane magnetic field while controlling the temperature of the sample to maintain a constant resistance near the superconducting transition. If, alternatively, the temperature is kept constant, hysteresis in seen in both the sheet resistance $R_s$ and the Hall resistance $R_{xy}$. The interplay of magnetic effects and superconducting state was later also probed by Ron et al. [54] in shadow-masked one dimensional nanowire with width ~50 nm, comparable to the superconducting coherence length. Ron reported a critical field $H_s$ (8mT to 12.5 mT) at the onset of a drop in the in-plane (both parallel and perpendicular to the current) magnetoresistance. Hysteresis loops in the magnetoresistance is reported for all directions with magnetization field close to $H_s$.

*Null results, inconsistencies and artifacts*

*Artifacts*

The first report of magnetism at the LaAlO$_3$/SrTiO$_3$ interface [4], in addition to reporting a Kondo resistance minimum (Figure 2a), also shows a hysteretic magnetoresistance curve that depends on the direction of the sweeping of the external field (Figure 7a) as well as the sweep rate. The hysteresis is stronger at lower temperatures or at higher sweep rate. After subsequent analyses, the authors determined that the hysteresis loops in those magnetoresistance curves can be ascribed to a magnetocaloric effect within the ceramic chip carrier used to hold the sample [55]. On the other hand, the other magnetic signature reported in [4], the Kondo resistance minimum, is not susceptible to this magnetocaloric effect.

*Null results*

While the 2DEG of LaAlO$_3$/SrTiO$_3$ is teeming with evidence of magnetism, there have been reports that are seemingly at odds, either quantitatively or qualitatively, with some of the primary reports. Using polarized neutron reflectometry, Fitzsimmons et al. [56] found essentially no evidence for magnetism in a variety of LaAlO$_3$/SrTiO$_3$ samples grown by two independent groups. The magnetization, which is calculated from the asymmetry in specular reflectivity as a function of wave vector transfer and neutron beam polarization (Figure 7c), is close to the signal-to-noise limit of the measurement.

*Scanning SQUID*

Follow-up scanning SQUID observations at the LaAlO$_3$/SrTiO$_3$ interface [18-20,28] by Wijnands [57] failed to reproduce observed magnetic patches (Figure 7b). The experiments were performed using a scanning SQUID microscope of similar design and sensitivity, with samples prepared under nominally identical conditions, as Bert et al. [18].

**Theories of magnetism at LaAlO₃/SrTiO₃ interface**

Theories of magnetism at the LaAlO₃/SrTiO₃ interface usually begin with a theory of the ferromagnetic phase. This phase was first attributed, by Pentcheva and Pickett [58], to the presence of oxygen vacancies at the interface that localize electrons in nearby Ti $3d$ states. Similar oxygen vacancies at the surface of SrTiO₃ may also introduce ferromagnetic order [59]. These localized states are located either in the Ti $d_{xy}$ orbitals [60,61], which lie very close to the interface and are thus easily localized by interface defects, or the Ti $e_g$ orbitals by a restructuring of the Ti $3d$ orbitals near oxygen vacancies [62-64]. In the former case the ferromagnetic state is assumed to arise from oxygen vacancies introducing localized magnetic moments. While in the latter case the localized $e_g$ electrons interact with itinerant electrons leading to Stoner magnetism.

With the discovery of localized ferromagnetic patches separated by a paramagnetic phase, additional detail was necessary. By taking the view that oxygen vacancies lead to an orbital reconstruction in nearby Ti, Pavlenko et al. [64] showed that ferromagnetism induced by oxygen vacancies would only occur above a critical vacancy density. This would then allow for the phase separation of superconductivity and ferromagnetism [65], which has alternatively been explained by a spiral-spin state in the $d_{xz}/d_{yz}$ bands by Banerjee et al. [66] while superconductivity is present in the $d_{xy}$ band. The patches [18-20,28] are possibly the broken spirals due to defects [66,67]. However, in this picture, the spiral states are made possible by the Rashba spin orbit coupling, whose magnitude can be controlled via electric field effect [36,39], the spiral states and, accordingly, ferromagnetic patches, are expected to be tunable via the electric field effect [66,67], but this is not observed by Bert et al. [28].

Alternative explanations of the coexistence of superconductivity and ferromagnetism rely upon a similar phase-separation picture. Michaeli et al. [61] argue that Fulde–Ferrell–Larkin–Ovchinnikov (FFLO) pairing can occur in the $d_{xz}/d_{yz}$ bands while the ferromagnetism arises from an Ruderman-Kittel-Kasuya-Yosida (RKKY) interaction amongst localized $d_{xy}$ electrons. Alternatively, Fidkowski et al. [68] argue that superconductivity in the interface is of a hybrid $s$- and $p$-wave pairing induced by the proximity effect with superconducting grains in the SrTiO$_3$ bulk while the ferromagnetism arises from Kondo interactions between localized moments and itinerant electrons, which the hybrid pairing is insensitive to.

In addition to the coexistence of ferromagnetism and superconductivity, there have also been several theoretical attemps to explain anisotropic magnetic effects. Fischer et al. [69] shows that Rashba spin-orbit coupling in a system with Stoner magnetism can result in an anisotropic spin susceptibility as well as a nematic phase arising from unequal occupation of the $d_{xz}/d_{yz}$ bands. Also using spin-orbit coupling, Fete et al. [12] presented a model where anisotropic magnetoconductance arises from an orbital reconstruction in which, at low carrier densities, the $d_{xy}$ band is responsible for the electron transport, and, as density increases, the $d_{xz}/d_{yz}$ bands begin to dominate and introduce anisotropic effects due to their nearly one-dimensional nature at the interface.

A recent attempt to explain many of these features simultaneously has been presented by Ruhman et al. [70]. In this theory, the magnetic features arise as a result of competition between Kondo screening in the $d_{xy}$ band and RKKY interactions in the $d_{xz}/d_{yz}$ bands. In addition to having paramagnetic (dominated by Kondo screening) and a ferromagnetic (dominated by localized magnetic moments/ordering of the RKKY interaction), there is also a spin-glass phase where the localized magnetic moments are insufficient to order the random RKKY interactions in the $d_{xz}/d_{yz}$ bands.

More recently, this many-body explanation of the magnetic effects in LaAlO$_3$/SrTiO$_3$ interfaces has been questioned by the temperature and density dependence of the giant magnetoresistance. In the case of Kondo screening, this effect should be very sensitive to thermalization effects, but this is not observed experimentally. Thus, Diez et al. [31] put forward a model based upon the quasi-classical Boltzmann diffusion equation with spin-orbit coupling and band anisotropy to explain this phenomenon as a single-particle effect.

**Main thesis: Two Distinct Classes of Magnetic Behavior in LaAlO$_3$/SrTiO$_3$**

Here we argue that the collection of experimental evidence for magnetism at the LaAlO$_3$/SrTiO$_3$ interface can be meaningfully sorted into two categories. The first is a *ferromagnetic* or Kondo liquid phase derived from local magnetic moments that are coupled via exchange interactions with itinerant electrons (Figure 9b). The ferromagnetic phase is favored when the interface is insulating, weakly conductive, or exhibiting locally insulating regions. The Kondo regime is favored when the density of itinerant electrons exceeds the density of local magnetic moments. We argue this is the magnetic phase that has been observed by scanning SQUID [18-20], SQUID [5,9], cantilever magnetometry [17], MFM [21,22], and MCD [71].

The second type of magnetism is a *metamagnetic* phase, resulting from an agglomeration of effects from the band structure of the *d* manifold of the Ti atoms in SrTiO$_3$, spin-orbit coupling, and electronic pairing. This second type of magnetism manifests itself in a variety of magnetotransport signatures including magnetoresistance anisotropy [5,7-12], sign changes of the magnetoresistance [15,31], and anomalous Hall effects [11,12,14,15]. Notably, these effects do not appear to interact with localized magnetic moments. Regarding interactions between ferromagnetism and superconductivity, we argue that the evidence suggests that there may indicate proximity effects but not necessarily true coexistence [72].

**Ferromagnetic and Kondo Regimes**

Ferromagnetism at the LaAlO$_3$/SrTiO$_3$ interface results from localized moments and their antiferromagnetic (RKKY) exchange with itinerant electrons [73]. The detail of the exchange interaction depends on the concentration of the itinerant carriers, as well as the concentration of localized moments. Variations involving superexchange [66] and double exchange have also been described theoretically [74,75]. As the local moments are antiferromagnetically exchange-coupled to itinerant electrons, ferromagnetic ordering between the local moments is expected [24,76] as long as the density of local moments $n_m$ exceeds that of the itinerant electrons $n_e$. The ferromagnetic order disappears as the interface become conductive. The dynamic magnetic screening of magnetization by itinerant carriers was directly observed by Bi et al. [21] using top-gated LaAlO$_3$/SrTiO$_3$ structures [21]. In the regime where $n_e > n_m$, we expect Kondo physics to dominate (Figure 9c).

The fact that ferromagnetism can be tuned electronically is crucial for reconciling many of the apparent inconsistencies—including null results—described earlier. For example, the experiments of Fitzsimmons [56] did not actively control electron density at the interface (Figure 10a) and may have taken place in the Kondo regime. Inconsistencies in observations of ferromagnetic patches—either the patches themselves or sample-to-sample variations—can possibly be reconciled in two ways. If a sample is locally insulating or simply has an excess of magnetic moments relative to the itinerant carrier density ($n_e < n_m$), then a ferromagnetic phase is favored. Local variations in the impurity density (or patches of magnetic moments) could fulfill this condition. Alternatively, local variations in the itinerant carrier density (with magnetic-impurity density presumably held constant) could produce a local ferromagnetic region. Surface adsorbates are known to strongly influence the local electron density [77-79].

The magnetic patches were not tuned within the backgate range explored (-70 V to 390 V) by Bert et al. [28], suggesting that the local density of the itinerant electrons $n_e$ and magnetic moments $n_m$ inside the magnetic patches must be substantially different from the rest of the sample and are therefore difficult to control via backgating, unlike the gate-tuned superfluid density for the rest of the sample. This explanation, however, does not seem to be compatible with the fact that no clear phase competition between the superfluid density and ferromagnetic patches is observed [80]. The independence of the magnetic effect on the backgate is also reported by Ron et al. [54] in shadow-mask created 1D nanowires. One possible explanation is that magnetic ordering exists at the intrinsically insulating boundary of the nanowire and harder to manipulated using electric field effect, if there are some ferromagnetic order. However, the low critical temperature (1 K) reported by Ron et al. [54] suggests a different origin or regime altogether.

*What are the local moments?*

Perhaps one of the outstanding questions relates to the physical origin of the local moments. They can either be extrinsic in nature, e.g., derived from the interface, or simply magnetic impurities. Alternatively, they can have an intrinsic origin, such as defect states from vacancies of composite atoms, cation substitution, or interdiffusion [81]. Here we discuss both possibilities and then narrow the discussion to the case for oxygen vacancies.

*Magnetic Impurities, Oxygen vacancies (vacancy clusters), and F-Centers*

Dilute magnetic impurities are generally found in SrTiO$_3$ substrates. Impurity levels on the order of one part per million have been reported [82-84], with large vendor-dependent variations [84]. For impurity concentrations of this level, it is possible to induce magnetism, as in the case of many dilute magnetic semiconductors [85]. However, observation of magnetic signals depends strongly on growth conditions, such as the oxygen partial pressure during growth of LaAlO$_3$ [9,86], post-growth annealing temperature and oxygen partial pressure [24,87]. This makes oxygen related defects such as vacancies and clusters the most common suspects. Both of these growth "knobs" can also tune the electron density. As mentioned, ferromagnetic hysteresis was only observed by Ariando [9] for samples with high oxygen pressure (P(O$_2$) = 10$^{-2}$ mbar), close to the metal-insulator transition. The more highly conductive samples grown at lower oxygen pressures did not show ferromagnetic hysteresis.

Oxygen vacancies are intrinsic, pervasive point defects for SrTiO$_3$. Oxygen vacancies and vacancy clusters modify the band structure and create deep, localized in-gap states [59,63], at, or in the vicinity of, defect sites. Each oxygen vacancy, in principle, donates 2 electrons. Depending on the configuration of vacancy clustering, these electrons can remain localized on the vacancy site, or fill the *d* orbitals of nearby Ti atoms. The defect sites with unpaired electrons are referred to as F-centers or color centers. Both F-centers and Ti 3d orbital can possess a local moment [88,89].

Strong evidence linking magnetic moments to oxygen vacancies comes from a report by Rice et al. [71] which used magnetic circular dichroism (MCD) to optically excite long-lived magnetic states of reduced SrTiO$_3$ single crystals. The sample was magnetized using a circularly polarized pump (Figure 11a) and probed via differential absorption of right- and left-circularly polarized light. The observed magnetization persists after the pumping beam is blocked, enabling long-lived magnetic states to be imprinted (Figure 11b). Samples doped with niobium did not show magnetization induced by MCD until oxygen vacancies were subsequently induced. The samples with oxygen vacancies did not exhibit magnetization after being re-oxygenated.

Following Rice et al. [71], Choi et al. [59] provided DFT calculations for in-gap states induced by oxygen vacancies of different valence, $V_O$, $V_O^+$, and $V_O^{2+}$, consistent with Rice et al [71]. Moments due to unpaired spins of Ti are only induced from $V_O$ and $V_O^+$. Furthermore, $V_O^{2+}$ dominate the population in the bulk, suggesting that magnetic moments may cluster around interfaces or ferroelastic domain boundaries. Calculations by Cuong et al. [90] suggest that vacancies tend to cluster into apical divacancy pairs, creating in-gap state of $e_g$ character. Altmeyer et al. [91] discuss magnetic moments of Ti $3d$ as a function of distance between the Ti atom and single oxygen vacancy or apical divacancies.

**Metamagnetism**

Here we describe a complementary set of effects, most of which are related to transport, that we label with a single term "metamagnetism". These effects generally take place well below room temperature (~35K) but above the superconducting transition. Many different effects take place within the same range of tunable parameters; we argue below that these effects are manifestations of the same underlying phenomena.

*Lifshitz transition*

Among all the reports from giant negative magnetoresistance, anisotropy in magnetoresistance and the anomalous Hall effect, the Lifshitz transition appears to play a central role. Its importance within the context of magnetotransport phenomena was first stressed by Joshua et al. [14]. The critical (Lifshitz) density (referred as $n_L$ in the following text) is where the $d_{xy}$-derived branches of the $t_{2g}$ bands are partially filled, and the $d_{xz}$/$d_{yz}$-derived bands begin to be populated (Figure 4). Giant negative magnetoresistance, magnetotransport anisotropy, and anomalous Hall effect are all observed only above the Lifshitz transition. The magnetoresistance near the Lifshitz transition evolves from being weak and weakly-dependent on angle to strong and highly anisotropic, with a crossover that depends on both magnetic field and electron density. The Lifshitz transition is also strongly linked to an observed tunable spin-orbit coupling [39] and takes place close to the maximum superconducting transition temperature [14].

*Anisotropic transport versus n, B*

As discussed previously, the anisotropy of the magnetoresistance undergoes a vivid transition from sinusoidal 2-fold oscillation (Figure 3b) to a 4-fold, or irregular oscillation (Figure 3c), when the carrier concentration $n_e$ is increased above the Lifshitz transition $n_L$. This transition between low and high anisotropy is also found to be a function of the external magnetic field (Figure 3d and e). The critical field for this transition depends on the carrier concentration, and a phase diagram (Figure 12b) was mapped out by Joshua et al. [15] showing that the critical magnetic field can vary from ~2 T to as large as 15 T as the carrier density is reduced to $n_L$. In separate measurements, Diez et al. [31] showed a large negative magnetoresistance that exhibits a strong back-gate dependence, becoming much more pronounced at large backgate values (Figure 6b). In a separate magnetotransport study of Ben Shalom et al. [36] (Figure 6a), a large negative magnetoresistance effect takes place at a critical magnetic field $\mu_0 H^*$ that ranges from 2 T to >18 T as the carrier density is reduced from $n = 7.8 \times 10^{13} \text{cm}^{-2}$ to $3.0 \times 10^{13} \text{cm}^{-2}$. It is worth noting that all of these experiments are probing a similar range of magnetic field and carrier density, and all of them are observing two distinct phases that are separated by a curve similar to what is shown in Figure 10b.

*Anomalous Hall effect*

As mentioned previously, there is also an critical carrier density $n_L$ as well as a carrier $n_e$ dependent critical in-plane $H_c^{\parallel}$ and out-of-plane field $H_c^{\perp}$, as mapped out by Joshua et al. [15]. The (model-independent) interpretation is that the offset and increase in Hall angle represents an excess magnetization that is "released" above a critical magnetic field, indicating a "gate-tunable polarized phase" [15]. Now the question becomes, what is this magnetizable degree of freedom, and what causes the sudden release as the field is increased? This magnetization does not behave as a single-component Langevin-type paramagentism, because it lacks the anticipated temperature dependence, i.e., *it is temperature-independent* [16]. As discussed, by adding an anomalous Hall term with Langevin-type paramagnetic field dependence, Gunkel et al. [16] was able to capture the small upturn in the Hall coefficient $R_H$ missed by the simple multiband model. However, Gunkel et al. [16] also pointed out that (i) a temperature-independent saturation field $B_C$ for the Hall coefficient $R_H$, and (ii) a temperature dependent $R_0^{AHE}$ deviates from the expected scaling of the Langevin-type spin-1/2 paramagnetic system.

*Correlated transport properties*

The anisotropic transport, anomalous Hall effect, and electron pairing without superconductivity all share features that change with carrier density and magnetic field in similar ways (see Figure 10b). Hence it is natural to consider the possibility that they are different manifestations of the same underlying phenomena. Here we discuss this possibility explicitly.

First, let us consider the AHE and its connection to electron pairing without superconductivity in a SET reported by Cheng et al. [49]. Based on the stability of the paired state conductance peaks with respect to magnetic field (region $|B| < B_P \approx 2$ T in Figure 8b) it is known that the electron pairs exist in a spin-singlet configuration, i.e., $S_{tot} = 0$. In the unpaired state (i.e., for $B > B_P$), electrons are now unpaired, with spin states $|S, S_z\rangle = |1/2, \pm 1/2\rangle$ that can subsequently Zeeman split and polarize in the excess magnetic field $B - B_P$. This polarization of unpaired spins should produce an anomalous Hall effect, similar to what is observed in bulk 2D experiments. The phase diagram for $H_C^{\parallel}$ from Joshua et al. [15] (Figure 12b) shows a monotonically decreasing dependence on the carrier density $n_e$. In the SET experiments, pairing fields as large as 5T were reported; in follow-up experiments on ballistic quantum channels the pairing transition has been observed at magnetic fields as large as 11 T [92]. The SET measurements only probe pairing at discrete values of carrier density, which itself is not directly measurable on the QD. For a single device, a few (2-5) data points are clearly observable. Figure 12a shows the pairing energy for a set of devices, plotted against a normalized gate voltage. While the scaling is not always monotonic, there is a general trend that favors larger pairing fields (in a particular device) with lower electron density. This behavior (Figure 12b) is similar to the crossover curve separating regions of large (anisotropic) and small (isotropic) magnetoresistance and anomalous Hall effect.

Regarding the transition between low and high anisotropy in transport, again, the collective evidence indicates that the paired state is most likely composed of $d_{xy}$ carriers, which, near the Lifshitz transition, is the preferred ground state due to a presumed attractive interaction shared only between $d_{xy}$ carriers. The $d_{xz}$ and $d_{yz}$ carriers, which have highly anisotropic band

23structure, become occupied when the pairing energy is overcome by sufficiently large magnetic fields. As the electron density increases above the Lifshitz point, the energy difference between the (unoccupied) $d_{xz}/d_{yz}$ states and the (occupied) $d_{xy}$ states reduces and eventually vanishes. That is to say, a monotonically decreasing pairing field is expected, regardless of the physical origin of the electron pairing itself. The crossover between attractive ($U < 0$) and repulsive ($U > 0$) regimes was observed by Cheng et al. [93] in transport measurements on devices similar to those depicted in Figure 8a but at higher carrier density.

Differences between the paired and unpaired phases are also manifested as negative magnetoresistance effects, notably the results from Ben Shalom et al. [36] (Figure 6a). In that experiment, a critical field $H^*$ was identified as the onset for giant negative magnetoresistance. The value of $H^*$ changes from <2T to >18T as the carrier density decreases toward what appears to be the Lifshitz transition (although about 2x higher than the value measured by Joshua et al. [15]). This behavior can also be incorporated into a framework in which changes in resistance may arise due to qualitative differences in transport between electron pairs and unpaired electrons. At sufficiently low carrier density, the electron pairs may propagate ballistically [94] or become localized.

**Future Directions**

*Empirical tests*

How would one test the proposed picture of two magnetisms? Regarding ferromagnetism and related phases, it is necessary to establish the origin of the magnetic moments. As discussed previously, oxygen vacancies are prime suspects for producing the required magnetic moments. Engineering oxygen vacancies via growth conditions and/or annealing can control the concentration of local moments [87], but it is not clear where exactly they are stabilized and whether they are single vacancies or clusters. To capture the physics, more detailed experimental and theoretical studies need to be performed. Concerning experiments that probe ferromagnetism, it is important to be able to control the concentration of itinerant electrons, a parameter that has been shown to be critical to the stabilization of ferromagnetic phases [21]. Top gating, as opposed to back gating through the $SrTiO_3$ substrate, is more effective in controlling carrier density and reaching a properly insulating phase. Finally, spatially-resolved measurements correlating oxygen vacancies and the magnetic moments, together with any long-range order they induce, could unequivocally establish the mechanism for dilute ferromagnetism at this interface.

### *Where are the oxygen vacancies?*

Here we turn specifically to the suspected role of oxygen vacancies. If they are indeed responsible for the magnetic moments, the question becomes: where exactly are these vacancies located? Are they pinned to the interfaces? Do they cluster around ferroelastic domain walls? While there is no direct experimental evidence showing that the oxygen vacancies are pinned within ferroelastic domain walls in $SrTiO_3$, in general, for perovskites, the movement of ferroelastic domain walls is restricted by defects, and could be easily pinned by them. Conversely, mobile oxygen vacancies may become trapped at ferroelastic domain walls. As pointed out by Goncalves-Ferreira et al. [95] for $CaTiO_3$ (Figure 11c), oxygen vacancies have lower energy when they are placed at a twin wall. Measurements using soundwaves suggest that domain walls are polar has been reported by Scott et al. [96].

Ferroelastic domains in $SrTiO_3$ have been imaged in a variety of ways, including optical microscopy [97], scanning SET (Figure 11d), scanning SQUID (Figure 11e), and low temperature SEM [98]. Ferroelastic domains are found to enhance the flow of the current [99], as well as the critical temperature of the superconductivity [100], and they play an important role in the gating of essentially all LaAlO$_3$/SrTiO$_3$ heterostructures at low temperatures. Scanning SQUID experiments, (Figure 11f) in which the probe gently touches the LaAlO$_3$ surface, have been demonstrated to produce significant changes in the magnitude and direction of ferromagnetic patches [20]. The force magnitude (~1 μN) is too small to create damage or otherwise depart from the elastic regime; however, it is plausible that this kind of force may result in a rearrangement of ferroelastic domains and therefore the spatial arrangement of magnetic moments. Further experimentation would be necessary to establish a clearer connection and to relate the ferroelastic domain structure to the presence of stabilized magnetic moments.

*What is the origin of electron pairing?*

In all of the experiments related to electron pairing thus far, the picture appears to be that there is a strongly attractive pairing interaction (Hubbard $U<0$) at low electron densities that is responsible for electron pairing and superconductivity. As the electron density increases, the sign of the electron-electron interaction changes ($U>0$) and superconductivity is ultimately suppressed. The paired, non-superconducting phase, close to the Lifshitz transition, becomes the dominant mode of transport, and yet it is far from being a Fermi liquid. Perhaps the biggest open question relates to the underlying physical mechanism for electron pairing. So far, there is no clear experiment (or theory) that favors a particular mechanism. A deeper understanding of electron pairing (and as a consequence superconductivity) in SrTiO$_3$-based systems would represent an important milestone in our overall ability to describe the physics of one of the richest material systems known.

**Acknowledgements**

We would like to acknowledge A. D. Caviglia, B. Kalisky, K. A. Moler, K. Nowack, and J. M. Triscone for helpful feedback on the manuscript. We thank NSF DMR (1124131, 1609519, 1124131) and ONR N00014-15-1-2847 for financial support.

**Figures and Figure Captions**

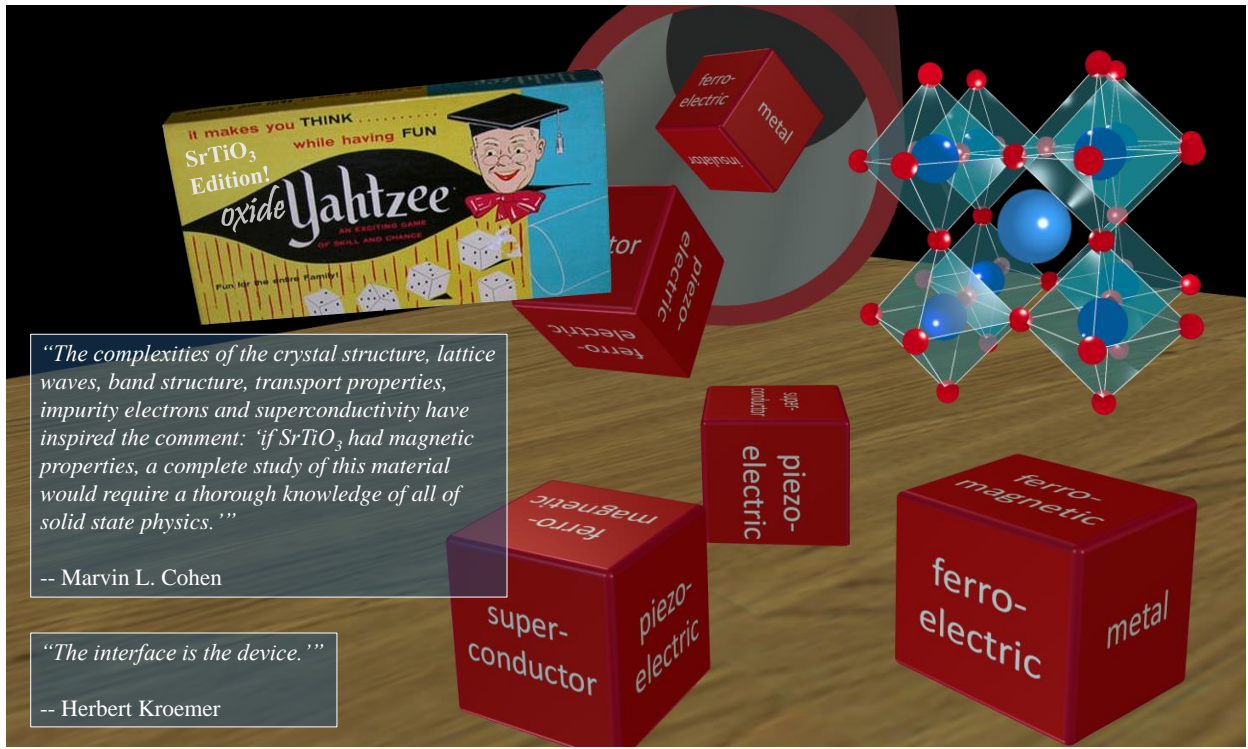

Figure 1. **Properties of LaAlO$_3$/SrTiO$_3$.** Properties that have been observed for SrTiO$_3$ from superconductivity via doping, ferroelectricity, piezoelectricity. Magnetism, however, is found at the interface of the LaAlO$_3$/SrTiO$_3$, even the neither of the two materials are magnetic. Quotes are from Ref. [1] and Ref. [101].

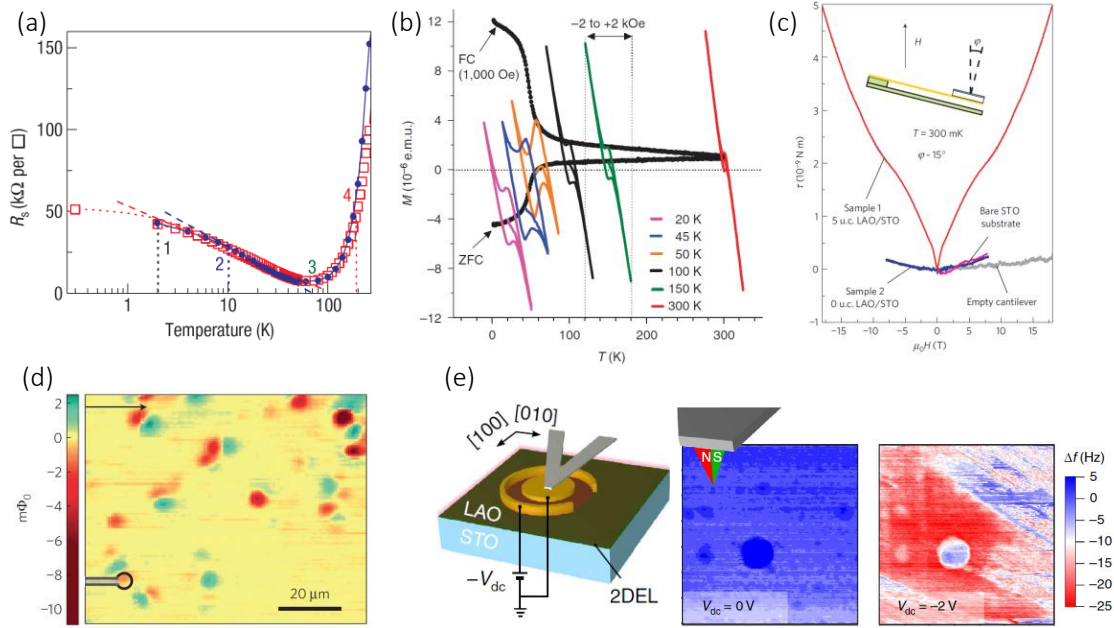

Figure 2. **Experimental evidence for ferromagnetism at the LaAlO$_3$/SrTiO$_3$ interface**. (a) Kondo resistance minimum (Adapted by permission from Macmillan Publishers Ltd: *Nature Materials* **6**, 493 - 496 (2007) [4], copyright 2007). (b) SQUID measurement: hysteresis loops taken at different temperature superimposed on top of the temperature dependence of magnetic moments (Adapted by permission from Macmillan Publishers Ltd: Nature Communications **2**, 188 (2010) [9], copyright 2010). (c) Cantilever-based magnetometry: the magnetic moment induced a torque under external magnetic field, the magnetization of the sample is inferred accordingly (Adapted by permission from Macmillan Publishers Ltd: *Nature Physics* **7**, 762–766 (2011) [17], copyright 2011). (d) With a micrometer-sized SQUID on a probe tip, microscopic magnetization can be imaged. Dipole-shaped patches are observed (Adapted by permission from Macmillan Publishers Ltd: *Nature Physics* **7**, 767–771 (2011) [18] copyright 2011). (e) Electrically-controlled ferromagnetism observed with magnetic force microscope. The magnetism signal is observed only when the interface is insulating (Adapted by permission from Macmillan Publishers Ltd: *Nature Communications* **5**, 5019 (2014) [21], copyright 2014).

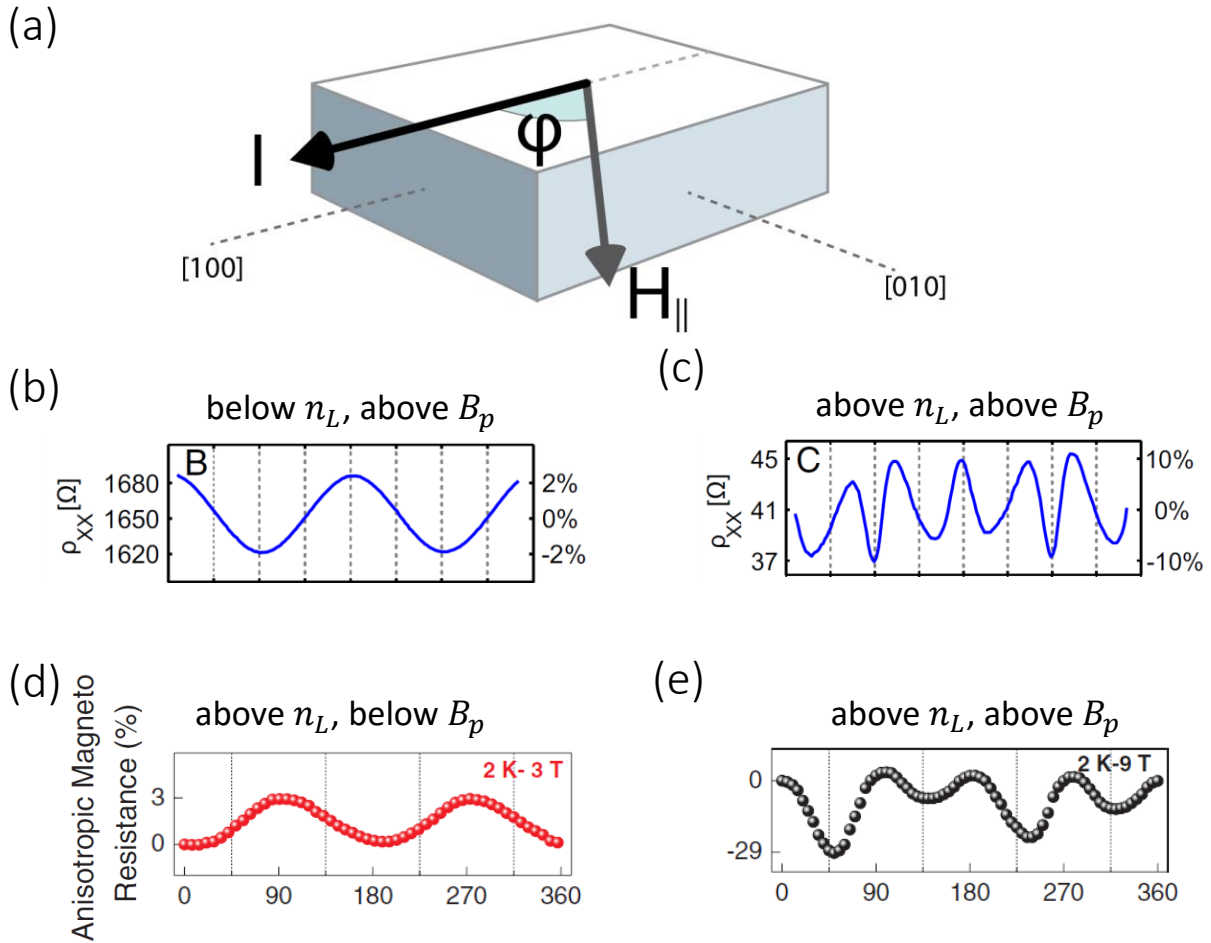

Figure 3. **Transport metamagnetism: anisotropic magnetoresistance.** (a) Setup for measuring in-plane anisotropy of the magnetoresistance and the metamagnetism transition critical field $B_p$. The anisotropy appears only when carrier concentration $n_e$ is greater than $n_L$ and the external magnetic field is greater than $B_p$ (which is also a function of $n_e$). A characteristic change in the magnetoresistance can be clearly seen from (b) to (c), when the carrier concentration is going above $n_L$, or (d) to (e), when the external field is exceeding $B_p$ ((b) and (c) adapted from Ref. [15]; (d) and (e) Reprinted figures with permission from Annadi et al., Phys. Rev. B **87**, 201102(R) (2013) [13]. Copyright 2013 by the American Physical Society).

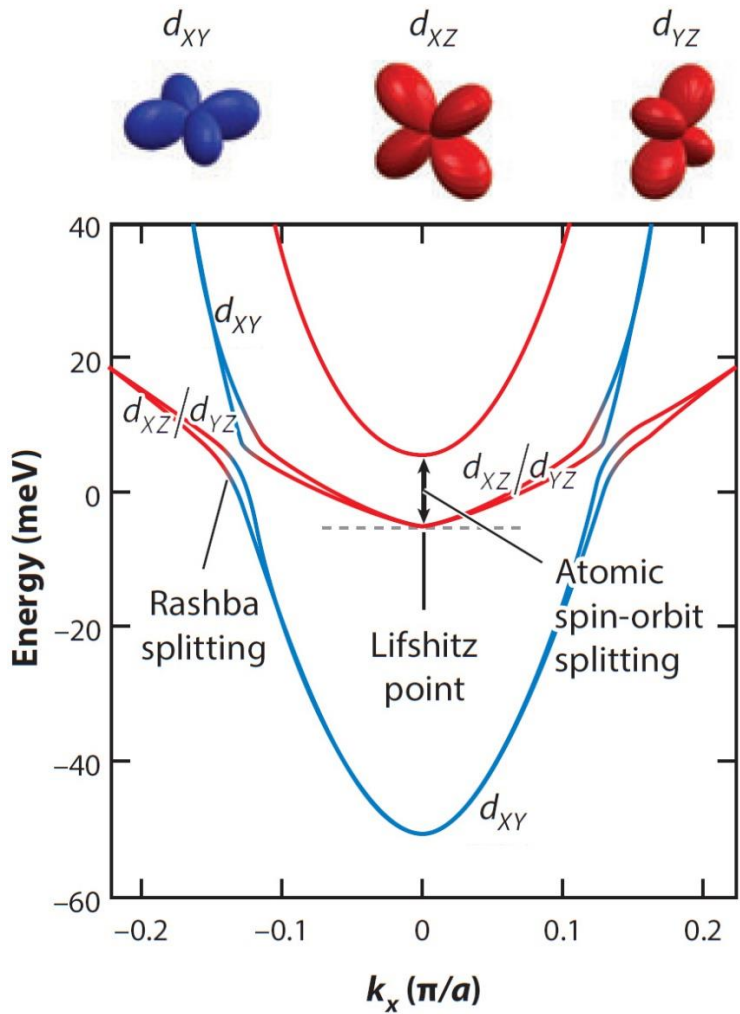

Figure 4. **Band structure of Ti 3d $t_{2g}$ orbitals**. The 3d orbitals of Ti split into a less energetic triplet $t_{2g}$ ($d_{xy}, d_{yz}, d_{xz}$) and a more energetic doublet $e_g$ ($d_{3z^2-r^2}, d_{x^2-y^2}$). $d_{xy}$ has the lowest energy of the three $t_{2g}$ orbitals, and is populated first. As the carrier concentration increases, $d_{yz}/d_{xz}$ start to be populated at Lifshitz point ($n_L$).

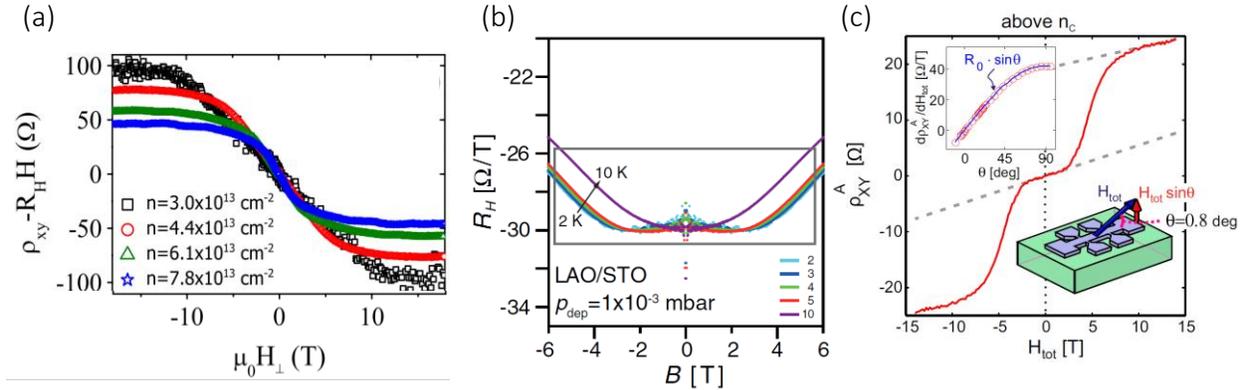

Figure 5. **Transport metamagnetism: nonlinear and anomalous Hall effect** at the LaAlO$_3$/SrTiO$_3$ interface. (a) Commonly reported nonlinearity in the Hall resistance: with standard Hall geometry and an out-of-plane external magnetic field, the magnitude of the Hall coefficient $R_H$ (the slope of the Hall resistance $R_{xy}$) decreases with the increased field and settles at some value smaller than the zero field $R_H$ (Reprinted figure with permission Ben Shalom et al., Phys. Rev. Lett. **104**, 126802 (2010) [36], Copyright 2010 by the American Physical Society). (b) Hall coefficient $R_H$ as function of applied field. While the overall parabolic-like shape can be explained by the multiband nature of SrTiO$_3$, the small upturn around zero field cannot. (adapted from [16]). (c) When applying a large in-plane field with a small out-of-plane field, the dependence of the Hall component on the Hall resistivity can be manipulated with the large in-plane field. There is a sudden release of the magnetic moments whenever the field exceeds a critical value (adapted from Ref ([15]).

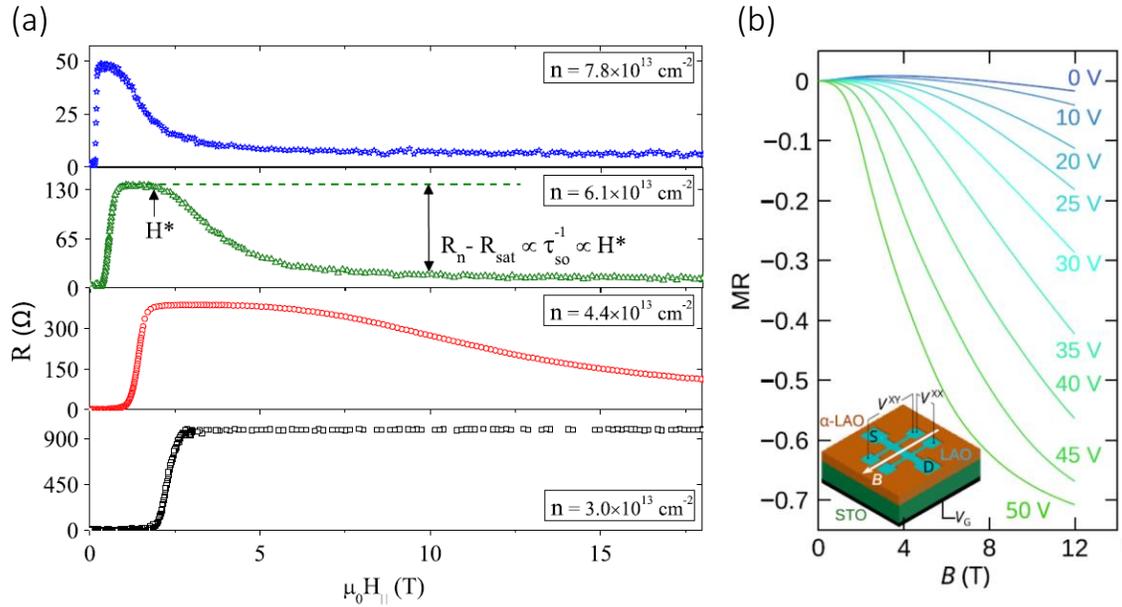

Figure 6. **Giant negative longitudinal ($R_{xx}$) magnetoresistance.** When an external magnetic field exceeds $H_c^{\parallel}$ and a carrier-concentration-dependent critical field $H^*$, the longitudinal magnetoresistance $R_{xx}$ drops significantly. (a) At $T = 20\ mK$, the evolution of the sheet resistance is plotted as a function of in-plane magnetic field (parallel to the current) at different carrier concentration. The resistance is restored at $H_c^{\parallel}$, followed by a field-independent interval from $H_c^{\parallel}$ to $H^*$, and then a sudden drop at $H^*$(Reprinted figure with permission from Ben Shalom et al., Phys. Rev. Lett. **104**, 126802 (2010) [31] Copyright 2010 by the American Physical Society). (b) At $T = 1.4K$ (above $T_c$), a drop of 70% in $R_{xx}$ is reported by Diez et al. (Reprinted figure with permission from Diez et al., Phys. Rev. Lett. **115**, 016803 (2015) [31] Copyright 2015 by the American Physical Society).

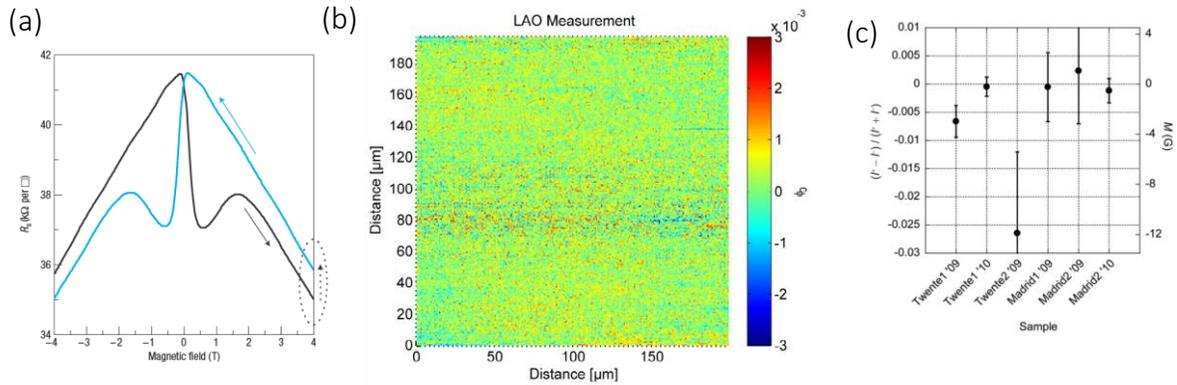

Figure 7. **Inconsistencies and null results.** Even though the signatures of magnetism are reported by many, there have also been controversies. (a) A hysteresis loop in the magnetoresistance previously thought as purely magnetic but later identified as magnetocaloric effect [55] (Adapted by permission from Macmillan Publishers Ltd: *Nature Materials* **6**, 493 - 496 (2007) [4], copyright 2007). (b) A null result reported by Wijnands et al.[57]: scanning SQUID on samples nominally equally to those reported in Bert et al. [18]. No ferromagnetic patches are found on the area of the sample explored (adapted with permission by Wijnands [57], copyright Wijnands 2014). (c) Using neutron spin-dependent reflectivity, where the magnetization is computed from asymmetry in the spin-dependent superlattice Bragg reflections as a function of wave vector transfer and neutron beam polarization, Fitzsimmons et al. [56] reported magnetization for $LaAlO_3$/$SrTiO_3$ superlattice close to the noise level through the presented 6 samples grown by two independent groups. (Reprinted figure with permission from Fitzsimmons et al., Phys. Rev. Lett. **107**, 217201 (2011) [56] Copyright (2011) by the American Physical Society).

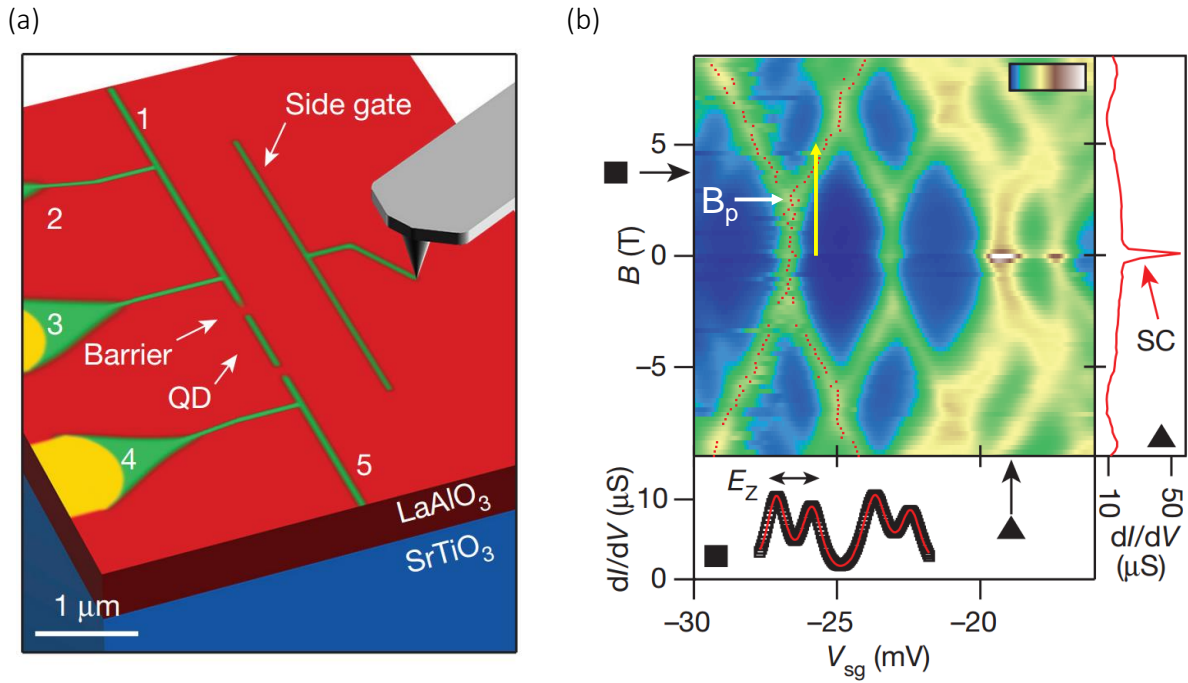

Figure 8. **Electron pairing without superconductivity.** Using a c-AFM lithography-defined single electron transistor, Cheng et al. [49] reported a phase of LaAlO$_3$/SrTiO$_3$ in which the electrons are paired but not condensed into the superconducting phase. (a) Schematics for the c-AFM lithography process and the single electron transistor device, consisting of a quantum dot (QD) and a sidegate. (Adapted from [49]}) (b) Evolution of the pairing transition as a function of external out-of-plane magnetic field and sidegate voltage (Adapted and annotated from [49] by permission from [49] the authors).

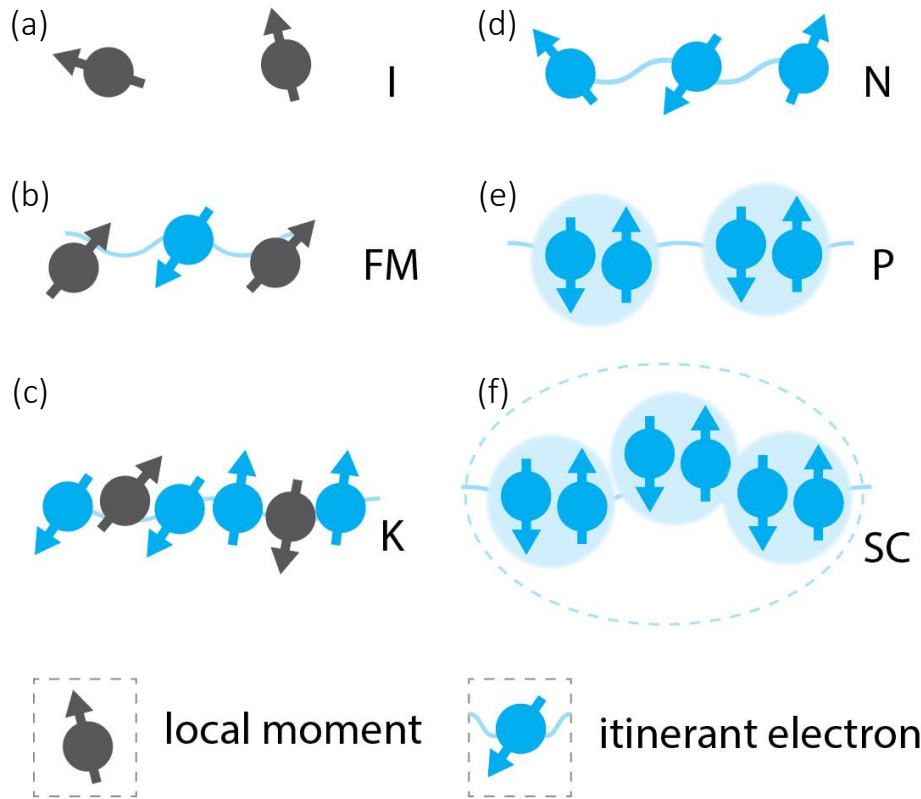

Figure 9. **Schematics for various phases for LaAlO$_3$/SrTiO$_3$.** (a) Insulating state (I): the concentration of itinerant electrons is not high enough to support conduction at the interface. (b) Ferromagnetic (FM): local moments are antiferromagnetically coupled to the itinerant electrons. When the concentration of the itinerant electrons is high enough to support long range ferromagnetic order but not too high to screen the local moments. (c) Kondo-screened (K) phase: when the concentration of itinerant electrons is further increased, the spins of itinerant electrons are coupled to localized moment. (d) Normal state (N): there are enough, when the itinerant electrons are enough to support conduction at the interface. (e) Paired state (P): where pairs of electrons are formed. (f) Superconducting (SC) phase.

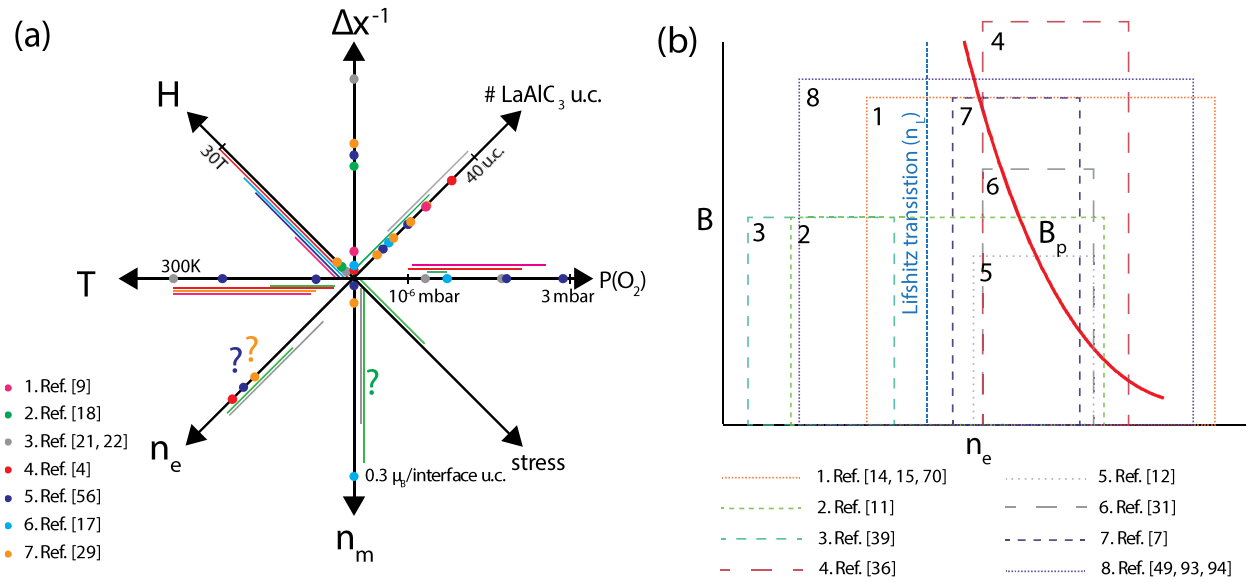

Figure 10. **Approximate parameter space and phase diagrams**. The parameter space explored by various research groups discussed in the main text. (a) Ferromagnetism. The knobs commonly tuned, but not limited to, are magnetic field (B), temperature (T), carrier concentration ($n_e$), moment concentration ($n_m$), mechanical stress, oxygen partial pressure for sample growth (P($O_2$)), number of LaAlO3 layers (# LaAlO3 u.c.), resolution ($\Delta x^{-1}$). We argue that the null results reported by Fitzsimmons et al. [56] could potentially be reconciled if the carrier concentration is controlled. (b) Metamagnetism parameter space explored by various works discussed in the text and the carrier concentration dependent critical field $B_p$ for the metamagnetic transition.

37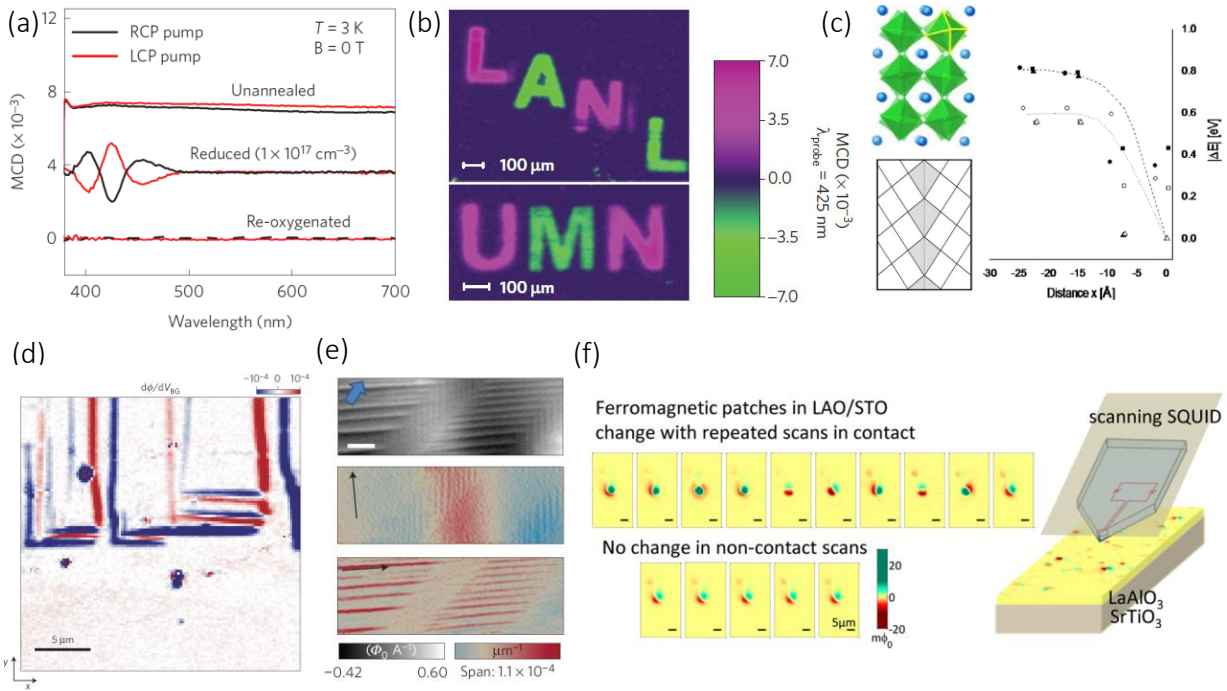

Figure 11. **Ferromagnetism, oxygen vacancies, and ferroelastic domains**. (a) Using magnetic circular dichroism (MCD), Rice et al. [71] reported a persistent magnetic signal for oxygen deficient $SrTiO_3$ samples. The signal disappears upon further re-oxygenation and reappears after introduction of oxygen vacancies (Adapted by permission from Macmillan Publishers Ltd: *Nature Materials* **13**, 481–487 (2014) [71] copyright 2014) (b) The persistent MCD signal created in real space (Adapted by permission from Macmillan Publishers Ltd: *Nature Materials* **13**, 481–487 (2014) [71] copyright 2014). (c) With DFT calculation for ferroelectric perovskite $BaTiO_3$, Goncalves-Ferreira et al. [95] found oxygen vacancies do have lower energy when around twin wall of the ferroelastic domains (Reprinted figures with permission from Goncalves-Ferreira et al., *Phys. Rev. B* **81**, 024109 (2010) [95]. Copyright (2010) by the American Physical Society.) (d) and (e) Experimentally observed ferroelastic domains using (d) a scanning SET (Adapted by permission from Macmillan Publishers Ltd: *Nature Materials* **12**, 1112–1118 (2013) [102] copyright 2013). and (e) scanning SQUID (Adapted by permission from Macmillan

Publishers Ltd: *Nature Materials* **12**, 1091–1095 (2013) [103] copyright 2013). (f) Mechanical manipulation of the magnetization using the probe of scanning SQUID (Adapted with permission from *Nano Lett.*, **2012**, 12 (8), pp 4055–4059 [20]. Copyright 2012 American Chemical Society).

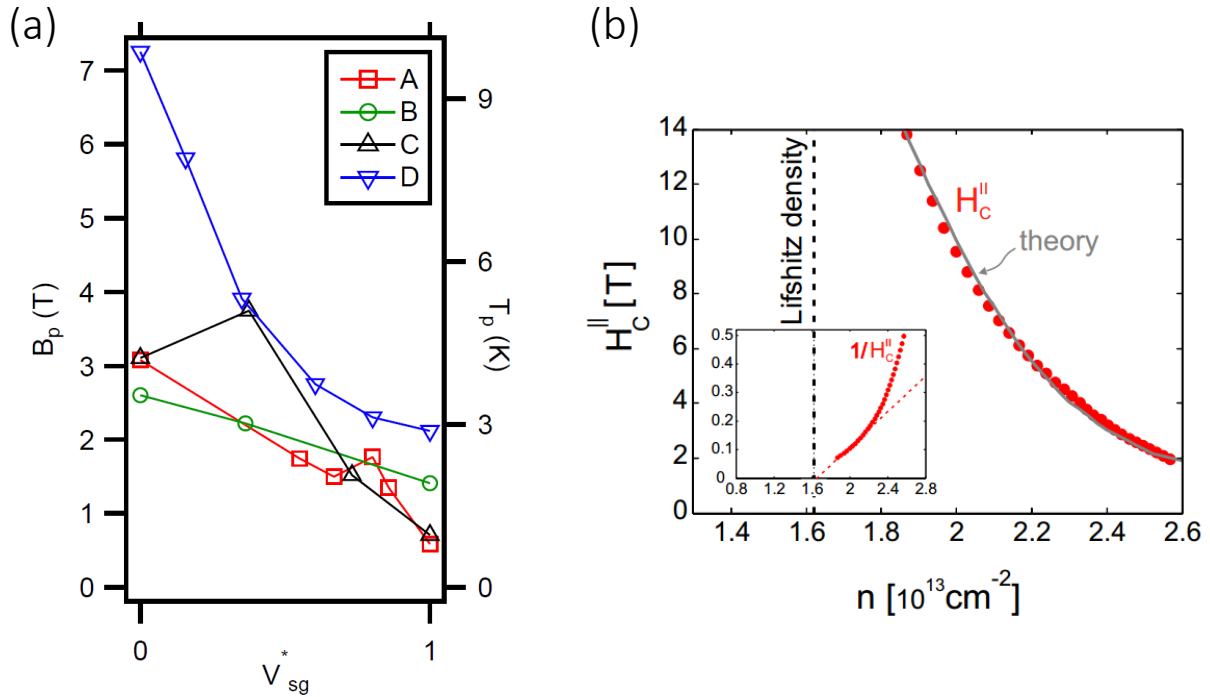

Figure 12. **Metamagnetism and pairing transition.** (a) Pairing field ($B_p$) as a function of voltage on sidegate ($V_{sg}$) (replotted from Ref [49]). (b) Density-dependent critical field $H_C^{\parallel}$ that marks the onset of AMR, AHE, and giant negative magnetoresistance (adapted from Ref. [15]).

**Table**

Table 1. Parameter space explored by the works discussed in the text.

| | max field (T) | temperature (K) | $n_e$ ($10^{13}$ cm$^{-2}$) | backgate voltage (V) | #LaAlO$_3$ unit cells (u.c.) | device size /dimensionality | P(O$_2$) (mbar) | Post Anneal? |
|---|---|---|---|---|---|---|---|---|
| Joshua et al. [15] | 14T | 50 mK, 2K, 4.2K | 1-3 | -50 to 450 | 6, 10 | mm/2D | $10^{-4}$ | o |
| Ben Shalom et al. [36] | 18T | 20 mK-100K | 0.1-5 | -50 to 50 | 8, 15 | mm/2D | $5 \times 10^{-4} - 10^{-3}$ | x |
| Caviglia et al. [39] | 12T | 30 mK - 20K | 1.85 - 2.2 | -340 to 320 | 12 | mm/2D | $6 \times 10^{-5}$, $10^{-4}$ | o |
| Fete et al. [12] | 7T | 1.5K | | | >4 | mm/2D | $10^{-4}$ | o |
| Ariando et al. [9] | 9T | 2K - 300K | | | 10 | mm/2D | $10^{-6} - 5 \times 10^{-2}$ | x |
| Ron et al. [54] Maniv et al. [104] | 18T | 20 mK, 60 mK | | -5 to 10 | 10, 6, 16, | 50 nm mm/2D | $10^{-4}$(Torr) | o |
| Cheng et al. [49] | 9T | 50 mK | | | 3.4 | nm/1D | $10^{-3}$ | o |

1    Cohen, M. L. Superconductivity in Low-Carrier-Density Systems: Degenerate Semiconductors. *Superconductivity: Part 1*, 615 (1969).
2    Bednorz, J. G. & Muller, K. A. Perovskite-type oxides - the new approach to High Tc superconductivity. *Nobel Lecture* (1987).
3    Ohtomo, A. & Hwang, H. Y. A high-mobility electron gas at the $LaAlO_3/SrTiO_3$ heterointerface. *Nature* **427**, 423-426 (2004).
4    Brinkman, A., Huijben, M., Van Zalk, M., Huijben, J., Zeitler, U., Maan, J. C., Van der Wiel, W. G., Rijnders, G., Blank, D. H. A. & Hilgenkamp, H. Magnetic effects at the interface between non-magnetic oxides. *Nature Materials* **6**, 493-496 (2007).
5    Hu, H.-L., Zeng, R., Pham, A., Tan, T. T., Chen, Z., Kong, C., Wang, D. & Li, S. Subtle Interplay between Localized Magnetic Moments and Itinerant Electrons in $LaAlO_3/SrTiO_3$ Heterostructures. *Acs Applied Materials & Interfaces*, 13630–13636 (2016).
6    Lee, M., Williams, J. R., Zhang, S., Frisbie, C. D. & Goldhaber-Gordon, D. Electrolyte Gate-Controlled Kondo Effect in $SrTiO_3$. *Physical Review Letters* **107**, 256601 (2011).
7    Ben Shalom, M., Tai, C. W., Lereah, Y., Sachs, M., Levy, E., Rakhmilevitch, D., Palevski, A. & Dagan, Y. Anisotropic magnetotransport at the $SrTiO_3/LaAlO_3$ interface. *Physical Review B* **80**, 140403 (2009).
8    Seri, S. & Klein, L. Antisymmetric magnetoresistance of the $SrTiO_3/LaAlO_3$ interface. *Physical Review B* **80**, 180410 (2009).
9    Ariando, Wang, X., Baskaran, G., Liu, Z. Q., Huijben, J., Yi, J. B., Annadi, A., Barman, A. R., Rusydi, A., Dhar, S., Feng, Y. P., Ding, J., Hilgenkamp, H. & Venkatesan, T. Electronic phase separation at the $LaAlO_3/SrTiO_3$ interface. *Nature Communications* **2**, 188 (2011).
10   Wang, X., Lu, W. M., Annadi, A., Liu, Z. Q., Gopinadhan, K., Dhar, S., Venkatesan, T. & Ariando. Magnetoresistance of two-dimensional and three-dimensional electron gas in $LaAlO_3/SrTiO_3$ heterostructures: Influence of magnetic ordering, interface scattering, and dimensionality. *Physical Review B* **84**, 075312 (2011).
11   Seri, S., Schultz, M. & Klein, L. Interplay between sheet resistance increase and magnetotransport properties in $LaAlO_3/SrTiO_3$. *Physical Review B* **86**, 085118 (2012).
12   Fête, A., Gariglio, S., Caviglia, A. D., Triscone, J. M. & Gabay, M. Rashba induced magnetoconductance oscillations in the $LaAlO_3$-$SrTiO_3$ heterostructure. *Physical Review B* **86**, 201105 (2012).
13   Annadi, A., Huang, Z., Gopinadhan, K., Wang, X. R., Srivastava, A., Liu, Z. Q., Ma, H. H., Sarkar, T. P., Venkatesan, T. & Ariando. Fourfold oscillation in anisotropic magnetoresistance and planar Hall effect at the $LaAlO_3/SrTiO_3$ heterointerfaces: Effect of carrier confinement and electric field on magnetic interactions. *Physical Review B* **87**, 201102 (2013).
14   Joshua, A., Pecker, S., Ruhman, J., Altman, E. & Ilani, S. A universal critical density underlying the physics of electrons at the $LaAlO_3/SrTiO_3$ interface. *Nature Communications* **3**, 1129 (2012).
15   Joshua, A., Ruhman, J., Pecker, S., Altman, E. & Ilani, S. Gate-tunable polarized phase of two-dimensional electrons at the $LaAlO_3/SrTiO_3$ interface. *Proceedings of the National Academy of Sciences* **110**, 9633 (2013).
16   Gunkel, F., Bell, C., Inoue, H., Kim, B., Swartz, A. G., Merz, T. A., Hikita, Y.,

Harashima, S., Sato, H. K., Minohara, M., Hoffmann-Eifert, S., Dittmann, R. & Hwang, H. Y. Defect Control of Conventional and Anomalous Electron Transport at Complex Oxide Interfaces. *Physical Review X* **6**, 031035 (2016).

17  Li, L., Richter, C., Mannhart, J. & Ashoori, R. C. Coexistence of magnetic order and two-dimensional superconductivity at LaAlO$_3$/SrTiO$_3$ interfaces. *Nature Physics* **7**, 762-766 (2011).

18  Bert, J. A., Kalisky, B., Bell, C., Kim, M., Hikita, Y., Hwang, H. Y. & Moler, K. A. Direct imaging of the coexistence of ferromagnetism and superconductivity at the LaAlO$_3$/SrTiO$_3$ interface. *Nature Physics* **7**, 767-771 (2011).

19  Kalisky, B., Bert, J. A., Klopfer, B. B., Bell, C., Sato, H. K., Hosoda, M., Hikita, Y., Hwang, H. Y. & Moler, K. A. Critical thickness for ferromagnetism in LaAlO$_3$/SrTiO$_3$ heterostructures. *Nature Communications* **3**, 922 (2012).

20  Kalisky, B., Bert, J. A., Bell, C., Xie, Y. W., Sato, H. K., Hosoda, M., Hikita, Y., Hwang, H. Y. & Moler, K. A. Scanning Probe Manipulation of Magnetism at the LaAlO$_3$/SrTiO$_3$ Heterointerface. *Nano Letters* **12**, 4055-4059 (2012).

21  Bi, F., Huang, M., Ryu, S., Lee, H., Bark, C.-W., Eom, C.-B., Irvin, P. & Levy, J. Room-temperature electronically-controlled ferromagnetism at the LaAlO$_3$/SrTiO$_3$ interface. *Nature Communications* **5**, 5019 (2014).

22  Bi, F., Huang, M., Lee, H., Eom, C.-B., Irvin, P. & Levy, J. LaAlO$_3$ thickness window for electronically controlled magnetism at LaAlO$_3$/SrTiO$_3$ heterointerfaces. *Applied Physics Letters* **107**, 082402 (2015).

23  Lee, J. S., Xie, Y. W., Sato, H. K., Bell, C., Hikita, Y., Hwang, H. Y. & Kao, C. C. Titanium d$_{xy}$ ferromagnetism at the LaAlO$_3$/SrTiO$_3$ interface. *Nature Materials* **12**, 703-706 (2013).

24  Salluzzo, M., Gariglio, S., Stornaiuolo, D., Sessi, V., Rusponi, S., Piamonteze, C., De Luca, G. M., Minola, M., Marré, D., Gadaleta, A., Brune, H., Nolting, F., Brookes, N. B. & Ghiringhelli, G. Origin of Interface Magnetism in BiMnO$_3$/SrTiO$_3$ and LaAlO$_3$/SrTiO$_3$ Heterostructures. *Physical Review Letters* **111**, 087204 (2013).

25  Liu, Z. Q., Lü, W. M., Lim, S. L., Qiu, X. P., Bao, N. N., Motapothula, M., Yi, J. B., Yang, M., Dhar, S., Venkatesan, T. & Ariando. Reversible room-temperature ferromagnetism in Nb-doped SrTiO$_3$ single crystals. *Physical Review B* **87**, 220405 (2013).

26  Huber, M. E., Koshnick, N. C., Bluhm, H., Archuleta, L. J., Azua, T., Björnsson, P. G., Gardner, B. W., Halloran, S. T., Lucero, E. A. & Moler, K. A. Gradiometric micro-SQUID susceptometer for scanning measurements of mesoscopic samples. *Review of Scientific Instruments* **79**, 053704 (2008).

27  Thiel, S., Hammerl, G., Schmehl, A., Schneider, C. W. & Mannhart, J. Tunable quasi-two-dimensional electron gases in oxide heterostructures. *Science* **313**, 1942-1945 (2006).

28  Bert, J. A., Nowack, K. C., Kalisky, B., Noad, H., Kirtley, J. R., Bell, C., Sato, H. K., Hosoda, M., Hikita, Y., Hwang, H. Y. & Moler, K. A. Gate-tuned superfluid density at the superconducting LaAlO$_3$/SrTiO$_3$ interface. *Physical Review B* **86**, 060503 (2012).

29  Salman, Z., Ofer, O., Radovic, M., Hao, H., Ben Shalom, M., Chow, K. H., Dagan, Y., Hossain, M. D., Levy, C. D. P., MacFarlane, W. A., Morris, G. M., Patthey, L., Pearson, M. R., Saadaoui, H., Schmitt, T., Wang, D. & Kiefl, R. F. Nature of Weak Magnetism in SrTiO$_3$/LaAlO$_3$ Multilayers. *Physical Review Letters* **109**, 257207 (2012).
30  Ben Shalom, M., Ron, A., Palevski, A. & Dagan, Y. Shubnikov-De Haas Oscillations in SrTiO$_3$/LaAlO$_3$ Interface. *Physical Review Letters* **105**, 206401 (2010).
31  Diez, M., Monteiro, A. M. R. V. L., Mattoni, G., Cobanera, E., Hyart, T., Mulazimoglu, E., Bovenzi, N., Beenakker, C. W. J. & Caviglia, A. D. Giant Negative Magnetoresistance Driven by Spin-Orbit Coupling at the LaAlO$_3$/SrTiO$_3$ Interface. *Physical Review Letters* **115**, 016803 (2015).
32  Stornaiuolo, D., Cantoni, C., De Luca, G. M., Di Capua, R., Di. Gennaro, E., Ghiringhelli, G., Jouault, B., Marre, D., Massarotti, D., Miletto Granozio, F., Pallecchi, I., Piamonteze, C., Rusponi, S., Tafuri, F. & Salluzzo, M. Tunable spin polarization and superconductivity in engineered oxide interfaces. *Nature Materials* **15**, 278-283 (2016).
33  Fête, A., Cancellieri, C., Li, D., Stornaiuolo, D., Caviglia, A. D., Gariglio, S. & Triscone, J.-M. Growth-induced electron mobility enhancement at the LaAlO$_3$/SrTiO$_3$ interface. *Applied Physics Letters* **106**, 051604 (2015).
34  Kim, J. S., Seo, S. S. A., Chisholm, M. F., Kremer, R. K., Habermeier, H. U., Keimer, B. & Lee, H. N. Nonlinear Hall effect and multichannel conduction in LaTiO$_3$/SrTiO$_3$ superlattices. *Physical Review B* **82**, 201407 (2010).
35  Gallagher, P., Lee, M., Petach, T. A., Stanwyck, S. W., Williams, J. R., Watanabe, K., Taniguchi, T. & Goldhaber-Gordon, D. A high-mobility electronic system at an electrolyte-gated oxide surface. *Nature Communications* **6**, 6437 (2015).
36  Ben Shalom, M., Sachs, M., Rakhmilevitch, D., Palevski, A. & Dagan, Y. Tuning spin-orbit coupling and superconductivity at the SrTiO$_3$/LaAlO$_3$ interface: a magnetotransport study. *Physical Review Letters* **104**, 126802 (2010).
37  Lee, Y., Clement, C., Hellerstedt, J., Kinney, J., Kinnischtzke, L., Leng, X., Snyder, S. D. & Goldman, A. M. Phase Diagram of Electrostatically Doped SrTiO$_3$. *Physical Review Letters* **106**, 136809 (2011).
38  Bell, C., Harashima, S., Kozuka, Y., Kim, M., Kim, B. G., Hikita, Y. & Hwang, H. Y. Dominant Mobility Modulation by the Electric Field Effect at the LaAlO$_3$/SrTiO$_3$ Interface. *Physical Review Letters* **103**, 226802 (2009).
39  Caviglia, A. D., Gabay, M., Gariglio, S., Reyren, N., Cancellieri, C. & Triscone, J. M. Tunable Rashba Spin-Orbit Interaction at Oxide Interfaces. *Physical Review Letters* **104**, 126803 (2010).
40  Narayanapillai, K., Gopinadhan, K., Qiu, X., Annadi, A., Ariando, Venkatesan, T. & Yang, H. Current-driven spin orbit field in LaAlO$_3$/SrTiO$_3$ heterostructures. *Applied Physics Letters* **105**, 162405 (2014).
41  Lesne, E., Fu, Y., Oyarzun, S., Rojas-Sanchez, J. C., Vaz, D. C., Naganuma, H., Sicoli, G., Attane, J. P., Jamet, M., Jacquet, E., George, J. M., Barthelemy, A., Jaffres, H., Fert, A., Bibes, M. & Vila, L. Highly efficient and tunable spin-to-charge conversion through Rashba coupling at oxide interfaces. *Nature Materials* **advance online publication** (2016).
42  Santander-Syro, A. F., Fortuna, F., Bareille, C., Rödel, T. C., Landolt, G., Plumb, N. C., Dil, J. H. & Radović, M. Giant spin splitting of the two-dimensional electron gas at the surface of SrTiO$_3$. *Nature Materials* **13**, 1085-1090 (2014).
43  McKeown Walker, S., Riccò, S., Bruno, F. Y., de la Torre, A., Tamai, A., Golias, E., Varykhalov, A., Marchenko, D., Hoesch, M., Bahramy, M. S., King, P. D. C., Sánchez-

Barriga, J. & Baumberger, F. Absence of giant spin splitting in the two-dimensional electron liquid at the surface of SrTiO$_3$ (001). *Physical Review B* **93**, 245143 (2016).
44  Garcia-Castro, A. C., Vergniory, M. G., Bousquet, E. & Romero, A. H. Spin texture induced by oxygen vacancies in strontium perovskite (001) surfaces: A theoretical comparison between SrTiO$_3$ and SrHfO$_3$ *Physical Review B* **93**, 045405 (2016).
45  Schooley, J. F., Hosler, W. R. & Cohen, M. L. Superconductivity in semiconducting SrTiO$_3$. *Physical Review Letters* **12**, 474-475 (1964).
46  Koonce, C. S., Cohen, M. L., Schooley, J. F., Hosler, W. R. & Pfeiffer, E. R. Superconducting Transition Temperatures of Semiconducting SrTiO$_3$. *Physical Review* **163**, 380-390 (1967).
47  Timusk, T. & Statt, B. The pseudogap in high-temperature superconductors: an experimental survey. *Reports on Progress in Physics* **62**, 61-122 (1999).
48  Richter, C., Boschker, H., Dietsche, W., Fillis-Tsirakis, E., Jany, R., Loder, F., Kourkoutis, L. F., Muller, D. A., Kirtley, J. R., Schneider, C. W. & Mannhart, J. Interface superconductor with gap behaviour like a high-temperature superconductor. *Nature* **502**, 528-531 (2013).
49  Cheng, G., Tomczyk, M., Lu, S. C., Veazey, J. P., Huang, M. C., Irvin, P., Ryu, S., Lee, H., Eom, C. B., Hellberg, C. S. & Levy, J. Electron pairing without superconductivity. *Nature* **521**, 196-199 (2015).
50  Kastner, M. A. The single-electron transistor. *Reviews of Modern Physics* **64**, 849-858 (1992).
51  Reyren, N., Thiel, S., Caviglia, A. D., Kourkoutis, L. F., Hammerl, G., Richter, C., Schneider, C. W., Kopp, T., Ruetschi, A. S., Jaccard, D., Gabay, M., Muller, D. A., Triscone, J. M. & Mannhart, J. Superconducting interfaces between insulating oxides. *Science* **317**, 1196-1199 (2007).
52  Caviglia, A. D., Gariglio, S., Reyren, N., Jaccard, D., Schneider, T., Gabay, M., Thiel, S., Hammerl, G., Mannhart, J. & Triscone, J. M. Electric field control of the LaAlO$_3$/SrTiO$_3$ interface ground state. *Nature* **456**, 624-627 (2008).
53  Dikin, D. A., Mehta, M., Bark, C. W., Folkman, C. M., Eom, C. B. & Chandrasekhar, V. Coexistence of Superconductivity and Ferromagnetism in Two Dimensions. *Physical Review Letters* **107**, 056802 (2011).
54  Ron, A., Maniv, E., Graf, D., Park, J. H. & Dagan, Y. Anomalous Magnetic Ground State in an LaAlO$_3$/SrTiO$_3$ Interface Probed by Transport through Nanowires. *Physical Review Letters* **113**, 216801 (2014).
55  Guduru, V. K. *Surprising Magnetotransport in Oxide Heterostructures*, Master's thesis, Radboud University, Nijmegen, the Netherlands., (2014).
56  Fitzsimmons, M. R., Hengartner, N. W., Singh, S., Zhernenkov, M., Bruno, F. Y., Santamaria, J., Brinkman, A., Huijben, M., Molegraaf, H. J. A., de la Venta, J. & Schuller, I. K. Upper Limit to Magnetism in LaAlO$_3$/SrTiO$_3$ Heterostructures. *Physical Review Letters* **107**, 217201 (2011).
57  Wijnands, T. *Scanning Superconducting Quantum Interference Device Microscopy*, Master's thesis, University of Twente, Enschede, the Netherlands. Retreived from: http://essay.utwente.nl/62800/1/Master_Thesis_Tom_Wijnands_openbaar.pdf, (2013).
58  Pentcheva, R. & Pickett, W. E. Charge localization or itinercy at LaAlO$_3$/SrTiO$_3$ interfaces: Hole polarons, oxygen vacancies, and mobile electrons. *Physical Review B* **74**,

035112 (2006).
59 Choi, H., Song, J. D., Lee, K.-R. & Kim, S. Correlated Visible-Light Absorption and Intrinsic Magnetism of SrTiO$_3$ Due to Oxygen Deficiency: Bulk or Surface Effect? *Inorganic Chemistry* **54**, 3759-3765 (2015).
60 Pentcheva, R. & Pickett, W. E. Ionic relaxation contribution to the electronic reconstruction at the n-type LaAlO$_3$/SrTiO$_3$ interface. *Physical Review B* **78**, 205106 (2008).
61 Michaeli, K., Potter, A. C. & Lee, P. A. Superconducting and Ferromagnetic Phases in SrTiO$_3$/LaAlO$_3$ Oxide Interface Structures: Possibility of Finite Momentum Pairing. *Physical Review Letters* **108**, 117003 (2012).
62 Chen, G. & Balents, L. Ferromagnetism in Itinerant Two-Dimensional $t_{2g}$ Systems. *Physical Review Letters* **110**, 206401 (2013).
63 Pavlenko, N., Kopp, T., Tsymbal, E. Y., Mannhart, J. & Sawatzky, G. A. Oxygen vacancies at titanate interfaces: Two-dimensional magnetism and orbital reconstruction. *Physical Review B* **86**, 064431 (2012).
64 Pavlenko, N., Kopp, T. & Mannhart, J. Emerging magnetism and electronic phase separation at titanate interfaces. *Physical Review B* **88**, 201104 (2013).
65 Pavlenko, N., Kopp, T., Tsymbal, E. Y., Sawatzky, G. A. & Mannhart, J. Magnetic and superconducting phases at the LaAlO$_3$/SrTiO$_3$ interface: The role of interfacial Ti 3d electrons. *Physical Review B* **85**, 020407(R) (2012).
66 Banerjee, S., Erten, O. & Randeria, M. Ferromagnetic exchange, spin-orbit coupling and spiral magnetism at the LaAlO$_3$/SrTiO$_3$ interface. *Nature Physics* **9**, 626-630 (2013).
67 Gabay, M. & Triscone, J.-M. Oxide heterostructures: Hund rules with a twist. *Nature Physics* **9**, 610-611 (2013).
68 Fidkowski, L., Jiang, H. C., Lutchyn, R. M. & Nayak, C. Magnetic and superconducting ordering in one-dimensional nanostructures at the LaAlO$_3$/SrTiO$_3$ interface. *Physical Review B* **87** (2013).
69 Fischer, M. H., Raghu, S. & Kim, E. A. Spin-orbit coupling in LaAlO$_3$/SrTiO$_3$ interfaces: magnetism and orbital ordering. *New Journal of Physics* **15**, 023022 (2013).
70 Ruhman, J., Joshua, A., Ilani, S. & Altman, E. Competition Between Kondo Screening and Magnetism at the LaAlO$_3$/SrTiO$_3$ Interface. *Physical Review B* **90**, 125123 (2013).
71 Rice, W. D., Ambwani, P., Bombeck, M., Thompson, J. D., Haugstad, G., Leighton, C. & Crooker, S. A. Persistent optically induced magnetism in oxygen-deficient strontium titanate. *Nature Materials* **13**, 481-487 (2014).
72 Cheng, G., Veazey, J. P., Irvin, P., Cen, C., Bogorin, D. F., Bi, F., Huang, M., Lu, S., Bark, C.-W., Ryu, S., Cho, K.-H., Eom, C.-B. & Levy, J. Anomalous Transport in Sketched Nanostructures at the LaAlO$_3$/SrTiO$_3$ Interface. *Physical Review X* **3**, 011021 (2013).
73 Ruderman, M. A. & Kittel, C. Indirect Exchange Coupling of Nuclear Magnetic Moments by Conduction Electrons. *Physical Review* **96**, 99-102 (1954).
74 Odkhuu, D., Rhim, S. H., Shin, D. & Park, N. La Displacement Driven Double-Exchange Like Mediation in Titanium dxy Ferromagnetism at the LaAlO$_3$/SrTiO$_3$. *Journal of the Physical Society of Japan* **85**, 043702 (2016).
75 Lechermann, F., Boehnke, L., Grieger, D. & Piefke, C. Electron correlation and magnetism at the LaAlO$_3$/SrTiO$_3$ interface: A DFT+DMFT investigation. *Physical*

*Review B* **90**, 085125 (2014).
76   Salluzzo, M. in *Oxide Thin Films, Multilayers, and Nanocomposites*   (eds Paolo Mele *et al.*)  181-211 (Springer International Publishing, 2015).
77   Xie, Y., Hikita, Y., Bell, C. & Hwang, H. Y. Control of electronic conduction at an oxide heterointerface using surface polar adsorbates. *Nature Communications* **2**, 494 (2011).
78   Brown, K. A., He, S., Eichelsdoerfer, D. J., Huang, M., Levy, I., Lee, H., Ryu, S., Irvin, P., Mendez-Arroyo, J., Eom, C.-B., Mirkin, C. A. & Levy, J. Giant conductivity switching of $LaAlO_3/SrTiO_3$ heterointerfaces governed by surface protonation. *Nature Communications* **7**, 10681 (2016).
79   Dai, W., Adhikari, S., Garcia-Castro, A. C., Romero, A. H., Lee, H., Lee, J.-W., Ryu, S., Eom, C.-B. & Cen, C. Tailoring $LaAlO_3/SrTiO_3$ Interface Metallicity by Oxygen Surface Adsorbates. *Nano Letters* **16**, 2739-2743 (2016).
80   Bert, J. A. *Superconductivity in Reduced Dimensions*, Doctoral thesis, Stanford University, CA, USA. Retreived from: http://www.stanford.edu/group/moler/theses/bert_thesis.pdf, (2012).
81   Salvinelli, G., Drera, G., Giampietri, A. & Sangaletti, L. Layer-Resolved Cation Diffusion and Stoichiometry at the $LaAlO_3/SrTiO_3$ Heterointerface Probed by X-ray Photoemission Experiments and Site Occupancy Modeling. *Acs Applied Materials & Interfaces* **7**, 25648-25657 (2015).
82   Coey, J. M. D. & Chambers, S. A. Oxide Dilute Magnetic Semiconductors—Fact or Fiction? *Mrs Bulletin* **33**, 1053-1058 (2008).
83   Garcia, M. A., Fernandez Pinel, E., de la Venta, J., Quesada, A., Bouzas, V., Fernandez, J. F., Romero, J. J., Martin-Gonzalez, M. S. & Costa-Kramer, J. L. Sources of experimental errors in the observation of nanoscale magnetism. *Journal of Applied Physics* **105**, 013925 (2009).
84   Coey, J. M. D., Venkatesan, M. & Stamenov, P. Surface magnetism of strontium titanate. *arXiv:1606.09422* (2016).
85   Dietl, T. & Ohno, H. Dilute ferromagnetic semiconductors: Physics and spintronic structures. *Reviews of Modern Physics* **86**, 187-251 (2014).
86   Huijben, M., Brinkman, A., Koster, G., Rijnders, G., Hilgenkamp, H. & Blank, D. H. A. Structure-Property Relation of $SrTiO_3/LaAlO_3$ Interfaces. *Advanced Materials* **21**, 1665-1677 (2009).
87   Liu, Z. Q., Sun, L., Huang, Z., Li, C. J., Zeng, S. W., Han, K., Lü, W. M., Venkatesan & Ariando. Dominant role of oxygen vacancies in electrical properties of unannealed $LaAlO_3/SrTiO_3$ interfaces. *Journal of Applied Physics* **115**, 054303 (2014).
88   Lopez-Bezanilla, A., Ganesh, P. & Littlewood, P. B. Magnetism and metal-insulator transition in oxygen-deficient $SrTiO_3$. *Physical Review B* **92**, 115112 (2015).
89   Lopez-Bezanilla, A., Ganesh, P. & Littlewood, P. B. Research Update: Plentiful magnetic moments in oxygen deficient $SrTiO_3$. *APL Mater.* **3**, 100701 (2015).
90   Cuong, D. D., Lee, B., Choi, K. M., Ahn, H.-S., Han, S. & Lee, J. Oxygen Vacancy Clustering and Electron Localization in Oxygen-Deficient $SrTiO_3$ : LDA+U Study. *Physical Review Letters* **98**, 115503 (2007).
91   Altmeyer, M., Jeschke, H. O., Hijano-Cubelos, O., Martins, C., Lechermann, F., Koepernik, K., Santander-Syro, A. F., Rozenberg, M. J., Valentí, R. & Gabay, M. Magnetism, Spin Texture, and In-Gap States: Atomic Specialization at the Surface of Oxygen-Deficient SrTiO$_3$. *Physical Review Letters* **116**, 157203 (2016).
92  Cheng et al, unpublished
93  Cheng, G., Tomczyk, M., Tacla, A. B., Lee, H., Lu, S., Veazey, J. P., Huang, M., Irvin, P., Ryu, S., Eom, C.-B., Daley, A., Pekker, D. & Levy, J. Tunable electron-electron interactions in LaAlO$_3$/SrTiO$_3$ nanostructures. *arXiv:1602.06029* (2016).
94  Tomczyk, M., Cheng, G., Lee, H., Lu, S., Annadi, A., Veazey, J. P., Huang, M., Irvin, P., Ryu, S., Eom, C.-B. & Levy, J. Micrometer-scale ballistic transport of electron pairs in LaAlO$_3$/SrTiO$_3$ nanowires. *Physical Review Letters* **117**, 096801 (2016).
95  Goncalves-Ferreira, L., Redfern, S. A. T., Artacho, E., Salje, E. & Lee, W. T. Trapping of oxygen vacancies in the twin walls of perovskite. *Physical Review B* **81**, 024109 (2010).
96  Scott, J. F., Salje, E. K. H. & Carpenter, M. A. Domain Wall Damping and Elastic Softening in SrTiO$_3$: Evidence for Polar Twin Walls. *Physical Review Letters* **109**, 187601 (2012).
97  Erlich, Z., Frenkel, Y., Drori, J., Shperber, Y., Bell, C., Sato, H. K., Hosoda, M., Xie, Y., Hikita, Y., Hwang, H. Y. & Kalisky, B. Optical Study of Tetragonal Domains in LaAlO$_3$/SrTiO$_3$. *Journal of Superconductivity and Novel Magnetism* **28**, 1017-1020 (2015).
98  Ma, H. J. H., Scharinger, S., Zeng, S. W., Kohlberger, D., Lange, M., Stöhr, A., Wang, X. R., Venkatesan, T., Kleiner, R., Scott, J. F., Coey, J. M. D., Koelle, D. & Ariando. Local Electrical Imaging of Tetragonal Domains and Field-Induced Ferroelectric Twin Walls in Conducting SrTiO$_3$. *Physical Review Letters* **116**, 257601 (2016).
99  Frenkel, Y., Haham, N., Shperber, Y., Bell, C., Xie, Y., Chen, Z., Hikita, Y., Hwang, H. Y. & Kalisky, B. Anisotropic Transport at the LaAlO$_3$/SrTiO$_3$ Interface Explained by Microscopic Imaging of Channel-Flow over SrTiO$_3$ Domains. *Acs Applied Materials & Interfaces* **8**, 12514-12519 (2016).
100 Hilary Noad, Eric M. Spanton, Katja C. Nowack, Hisashi Inoue, Minu Kim, Tyler A. Merz, Christopher Bell, Yasuyuki Hikita, Ruqing Xu, Wenjun Liu, Arturas Vailionis, Harold Y. Hwang & Moler, K. A. Enhanced Superconducting Transition Temperature due to Tetragonal Domains in Two-Dimensionally Doped SrTiO$_3$. *arXiv:1605.08418* (2016).
101 Kroemer, H. Nobel Lecture: Quasielectric fields and band offsets: teaching electrons new tricks. *Reviews of Modern Physics* **73**, 783-793 (2001).
102 Honig, M., Sulpizio, J. A., Drori, J., Joshua, A., Zeldov, E. & Ilani, S. Local electrostatic imaging of striped domain order in LaAlO$_3$/SrTiO$_3$. *Nature Materials* **12**, 1112-1118 (2013).
103 Kalisky, B., Spanton, E. M., Noad, H., Kirtley, J. R., Nowack, K. C., Bell, C., Sato, H. K., Hosoda, M., Xie, Y., Hikita, Y., Woltmann, C., Pfanzelt, G., Jany, R., Richter, C., Hwang, H. Y., Mannhart, J. & Moler, K. A. Locally enhanced conductivity due to the tetragonal domain structure in LaAlO$_3$/SrTiO$_3$ heterointerfaces. *Nature Materials* **12**, 1091-1095 (2013).
104 Maniv, E., Shalom, M. B., Ron, A., Mograbi, M., Palevski, A., Goldstein, M. & Dagan, Y. Strong correlations elucidate the electronic structure and phase diagram of LaAlO$_3$/SrTiO$_3$ interface. *Nature Communications* **6**, 8239 (2015).